\documentclass[pra,aps,amsmath,amssymb,amsfonts,twocolumn,nofootinbib,longbibliography,superscriptaddress,floatfix]{revtex4-1}
\usepackage{amssymb}
\usepackage{bm,mathrsfs}
\usepackage{graphicx}
\usepackage{epsfig}
\usepackage{amsmath,bbm}
\usepackage{amsfonts,amssymb}
\usepackage{times}
\usepackage[mathscr]{eucal}
\usepackage{verbatim}
\usepackage[sort&compress]{natbib}
\usepackage{amsmath}
\usepackage{bm}
\usepackage[colorlinks,breaklinks,linkcolor=blue,anchorcolor=blue,citecolor=blue,urlcolor=blue]{hyperref}

\begin{document}

\title{Single temporal-pulse-modulated parameterized controlled-phase gate for Rydberg atoms}

\author{X. X. Li}
\affiliation{Center for Quantum Sciences and School of Physics, Northeast Normal University, Changchun 130024,  China}

\author{X. Q. Shao}
\email{shaoxq644@nenu.edu.cn}
\affiliation{Center for Quantum Sciences and School of Physics, Northeast Normal University, Changchun 130024, China}
\affiliation{Center for Advanced Optoelectronic Functional Materials Research, and Key Laboratory for UV Light-Emitting Materials and Technology
of Ministry of Education, Northeast Normal University, Changchun 130024, China}

\author{Weibin Li}
\email{weibin.Li@nottingham.ac.uk}
\affiliation{School of Physics and Astronomy, The University of Nottingham, Nottingham NG7 2RD, United Kingdom}
\begin{abstract}
We propose an adiabatic protocol for implementing a controlled-phase gate CZ$_{\theta}$ with continuous $\theta$ of neutral atoms through a symmetrical two-photon excitation process via the second resonance line, $6P$ in $^{87}$Rb, with a single-temporal-modulation-coupling of the ground state and intermediate state.
Relying on different adiabatic paths, the phase factor $\theta$ of CZ$_{\theta}$ gate can be accumulated on the logic qubit state $|11\rangle$ alone by calibrating the shape of the temporal pulse where strict zero amplitudes at the start and end of the pulse are not needed.  For a wide range of $\theta$, we can  obtain the fidelity of CZ$_{\theta}$ gate over $99.7\%$ in less than $1~\mu$s, in the presence of spontaneous emission from intermediate and Rydberg states. And in particular for $\theta=\pi$, we benchmark the performance of the CZ gate by taking into account various experimental imperfections, such as Doppler shifts,  fluctuation of Rydberg-Rydberg
interaction strength, inhomogeneous Rabi frequency, and noise of driving fields, etc,
and show that the predicted fidelity is able to maintain at about $98.4\%$ after correcting the measurement error. This gate protocol provides a robustness against the fluctuation of pulse amplitude and a flexible way for adjusting the entangling phase, which may contribute to the experimental implementation of near-term noisy intermediate-scale quantum (NISQ)
 computation and algorithm with neutral-atom systems.

\end{abstract}

\maketitle

\section{introduction}

Two-qubit entangling gates, such as the controlled-Z (CZ) gate and the equivalent controlled-NOT (CNOT) gate, are at the center of universal quantum computation \cite{RevModPhys.74.347,Ladd2010,Wendin2017}. A challenge is to realize fast and high-fidelity gate protocol experimentally, where quantum logic gates have been embodied in various physical systems, such as nuclear magnetic resonance (NMR), quantum dots, ion traps, semiconductor silicon, and Josephson junction \cite{PhysRevB.100.035304,PhysRevB.103.235314,PhysRevLett.117.060504,PhysRevLett.117.060505,Barends2014,PhysRevApplied.15.064005,PhysRevLett.123.120502}. Among many physical systems, neutral atoms have also been considered for the realization of logic gates, due to the long-lived encoding in atomic hyperfine states and the possibility of manipulating and measuring qubit states via laser lights \cite{Bloch2008,RevModPhys.82.2313}. Recently coherent operation of internal and external states of Rydberg atoms has become possible. Combined with the strong interaction between Rydberg atoms (e.g. the blockade effect) \cite{Urban2009,PhysRevLett.93.063001,Gatan2009}, fast quantum logic gates have been realized with Rydberg atoms \cite{PhysRevLett.85.2208,PhysRevLett.87.037901,Brion2007,PhysRevA.82.034307,PhysRevA.84.042315,PhysRevA.88.062337,PhysRevLett.113.123003,PhysRevA.92.030303,Saffman_2016,PhysRevA.93.012306,PhysRevA.95.022319,PhysRevApplied.9.051001,PhysRevA.95.043429,PhysRevApplied.7.064017,PhysRevA.98.052324,PhysRevA.98.032306,PhysRevA.98.062338,PhysRevApplied.11.044035,Yin:20,LiRui2021,PhysRevA.105.032417}. Zhang {\it et al}. analyzed the gates errors under the Rydberg blockade in detail and pointed out that the errors cannot be made arbitrarily small by addressing higher-lying Rydberg levels for the diminution of the blockade effect \cite{PhysRevA.85.042310}. So far, the theoretical limit has been extended to $F> 0.9999$ \cite{PhysRevA.94.032306,PhysRevA.96.042306}, while experiments have not reached such fidelity \cite{PhysRevLett.104.010503,PhysRevLett.104.010502,PhysRevA.82.030306,PhysRevA.92.022336,PhysRevLett.123.230501}.

One of the reasons that leads to the gap between experiments and theoretical prediction is the unexpectedly large loss of atoms remaining in the Rydberg state in the gate operation. By comparing the standard Rydberg blockade CZ pulse sequence \cite{PhysRevLett.85.2208} with the protocols with continuing pulses on the ground-Rydberg transition \cite{PhysRevLett.123.230501,PhysRevA.92.022336}, it is shown that the continuous pumping protocols are more advantageous. In recent reports, Levine {\it et al}. experimentally implemented the multiqubit gates in one-dimensional geometry, in which the fidelity of CZ gate $F > 0.97$ \cite{PhysRevLett.123.170503}, while Graham {\it et al}. realized CZ gate with $F = 0.89$ in a two-dimensional qubit array \cite{PhysRevLett.123.230501}. Moreover, via Single-modulated-pulse off-resonant modulated driving (SORMD) embedded in two-photon transition for Rb atoms within the Rydberg blockade region, Fu {\it et al}. also experimentally realized a CZ gate with $F = 0.980(7)$ after correcting the state preparation and measurement (SPAM) errors \cite{PhysRevA.105.042430}.

On the other hand, a new class of algorithms for quantum computing is based on adiabatic evolution, which provides a strategy for suppressing certain error mechanisms such as atomic motions \cite{Science2001}. For this reason, it is a promising method to realize a robust high-fidelity logic gate
\cite{PhysRevLett.100.170504,Muller2011,PhysRevA.90.032329,PhysRevA.90.033408,PhysRevA.89.032334,PhysRevA.89.030301,PhysRevA.91.032304,PhysRevA.94.062307,PhysRevA.96.022321,PhysRevA.97.032701,Zhang2020,PhysRevA.101.062309,PhysRevA.101.030301,PhysRevA.103.062607}. The protocols with adiabatic rapid passage (ARP) pulses show that the gate fidelity can reach over 0.999 \cite{PhysRevA.94.062307,PhysRevA.101.030301,PhysRevA.101.062309}. However, the difficulty to apply this to Rydberg atom gates is that large energies are required to directly excite atoms from the ground state to Rydberg state. For example, the corresponding laser wavelength in the single-photon excitation of $^{87}$Rb atoms is in the ultraviolet region, which is difficult to produce and use. Therefore, the Rydberg excitation is generally realized by a two-photon excitation scheme. Most adiabatic protocols use STIRAP pulse sequences, in which both two laser beams are temporal-modulated, and some even combine a time modulated detuning, which is complicated in practice. Of which, a ``STIRAP-inspired" gate with globally optimized pulses Saffman {\it et al}. provided can reach higher fidelity $F = 0.997$ \cite{PhysRevA.101.062309}. However, this pulse sequence is sensitive to the fluctuation of excitation laser intensity.

\begin{figure*}
\centering\scalebox{0.38}{\includegraphics{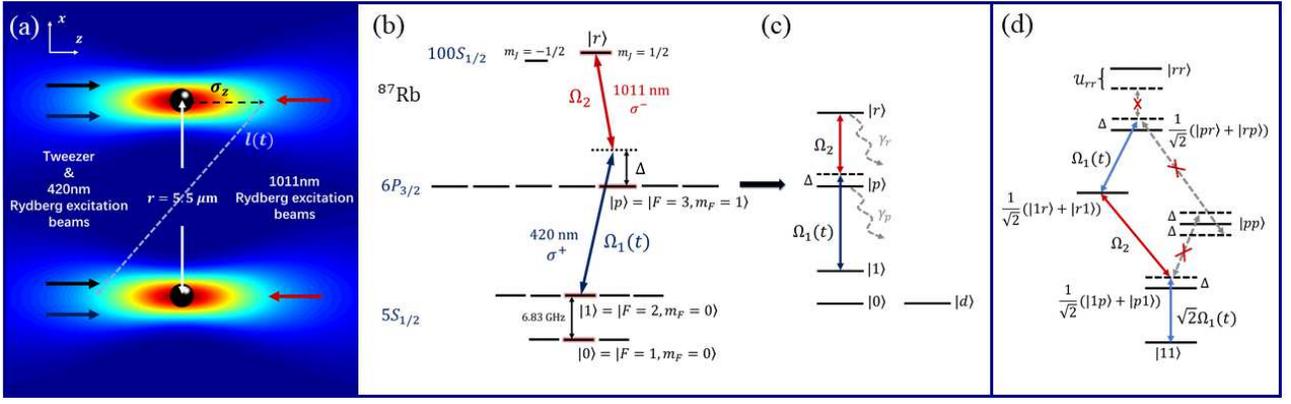}}
\caption{\label{ple1}(a) Experimental geometry. Two single atoms are trapped in two tweezers separated by about $5.5~\mu$m with tweezer beam and Rydberg excitation beams propagating along quantized $z$-axis.
(b) Relevant levels of $^{87}$Rb. The $5S_{1/2}$ hyperfine clock states $|0\rangle\equiv|F=1, m_{F}=0\rangle$, $|1\rangle\equiv|F=2, m_{F}=0\rangle$ are chosen as two ground states. To excite the Rydberg state we use a two-photon scheme with wavelengths of $420$~nm and $1011$~nm.
(c) Equivalent energy-level configuration of neutral atom qubit. Level $|d\rangle$ is an uncoupled state representing the leakage levels outside qubit basis $\{|0\rangle, |1\rangle\}$.
(d) The effective system dynamics initial from state $|11\rangle$, where ${\cal U}_{rr}$ is the vdW interaction between Rydberg states.}
\end{figure*}

It is worth noting that compared with the standard CZ gate, a parameterized controlled-phase  ($\textmd{CZ}_{\theta}$)  gate with flexible angle adjustment plays an important role in implementing quantum algorithms. Especially for applying the quantum approximate optimization algorithm (QAOA) to solve the combinatorial
problem that can be mapped onto finding the ground state of an Ising Hamiltonian, the application of $\textmd{CZ}_{\theta}$ gate will greatly simplify the synthesis of quantum circuits and improve the success probability and performance as the number of QAOA layers increase \cite{PhysRevApplied.14.034010,PRXQuantum.1.020304,Graham2022}.

Therefore, in viewing these practical challenges, we propose a new adiabatic method to realize a continuous controlled-phase gate set in neutral-atom system.
By symmetrically driving qubit atoms with a single-modulated pulse of blue detuned to the transition between ground state and the intermediate state, and a constant-amplitude
pulse that is red detuned to the transition between the intermediate state and the excited Rydberg state, we can acquire an arbitrary dynamical phase factor of $\theta\in[0.08\pi, \pi]$ accumulated on logic qubit state $|11\rangle$ alone within the Rydberg blockade regime by simply modulating the shape of the temporal pulse.
The prominent advantages of our scheme are threefold:
(i) The temporal pulse can be adopted as a Gaussian pulse or any other pulses with no need of a strict zero amplitude at the start and the end,  but small enough to ensure that the adiabatic condition is established.
(ii) For a wide range of $\theta$, we can still obtain the $\textmd{CZ}_{\theta}$ gate fidelity over $99.7\%$ in less than $1~\mu$s, even considering spontaneous dissipation at room temperature.
(iii) As a specific case of $\theta=\pi$, we assess the performance of the CZ gate by considering the technical imperfections in experiment, and find the predicted fidelity is able to maintain at about $98.4\%$ for a realistic situation after correcting the detection errors, which may be helpful to the experimental implementation of quantum computation and quantum simulation in the neutral-atoms system.

The remainder of the paper is organized as follows.  In Sec.~\ref{sec2}, we introduce the basic principle of the scheme and analytically show how the fast and high-fidelity parameterized controlled-phase gate is adiabatically constructed.
In Sec.~\ref{secnew}, we take the Max-Cut issue and the Fourier transform as two examples to demonstrate the benefits of using the  CZ$_\theta$ gate in quantum computing as opposed to the traditional CZ gate.
In Sec.~\ref{sec3}, we take the CZ gate as an example and discuss in detail the experimental feasibility and the gate errors introduced by technical imperfections, e.g. the Doppler shifts, the fluctuation of Rydberg-Rydberg interaction strengths, the inhomogeneous Rabi frequency, the fluctuation and noise of external fields, and the detection errors, and make a comparison with previous works in the literature.
In Sec.~\ref{sec4}, we give two examples of realizing the controlled-phase gate with non-Gaussian temporal pulses.
In Sec.~\ref{sec5}, we briefly discuss the application of the proposed scheme to cesium atoms, and obtain that the fidelity of the CZ gate can be achieved $99.81\%$ by  fully taking into account the spontaneous emission from intermediate and Rydberg states. Finally, we make a conclusion.

\section{parameterized controlled-phase gate}\label{sec2}
The parameterized controlled-phase gate ($\textmd{CZ}_{\theta}$) is a two-qubit gate belonging to controlled unitary operations. it can pick up a phase $\theta$ on the target state $|1\rangle$ if and only if the control qubit is in state $|1\rangle$ \cite{PhysRevLett.74.4083}. In the computational basis $\{|00\rangle, |01\rangle, |10\rangle, |11\rangle\}$, it can be defined as the unitary transformation
\begin{equation}
U_{\textmd{CZ}_{\theta}}=\left[
 \begin{array}{cccc}
 1 & 0 & 0 & 0 \\
 0 & 1 & 0 & 0 \\
 0 & 0 & 1 & 0 \\
 0 & 0 & 0 & e^{-i\theta} \\
 \end{array}
\right].
\end{equation}

The physical system considered to realize this operation is a pair of $^{87}$Rb atoms trapped in two tweezers with separation $r$ shorter than the blocking radius, as shown in Fig.~\ref{ple1}(a). The relevant levels are displayed in Fig.~\ref{ple1}(b). The logic qubit is encoded on  $|0\rangle\equiv|F=1, m_{F}=0\rangle$ and $|1\rangle\equiv|F=2, m_{F}=0\rangle$ of $5S_{1/2}$ hyperfine clock states with splitting $2\pi\times 6.83$~GHz, and the Rydberg state $|r\rangle\equiv|100S_{1/2}, m_{j}=1/2\rangle$ is used to mediate the interaction between atoms.
To coherently drive atoms from ground states to the Rydberg states, we apply two-photon excitation lasers,
a $\sigma_{+}$ polarized $420~$nm laser and a $\sigma_{-}$ polarized $1011~$nm laser, via the second resonance line $|p\rangle\equiv|6p_{3/2}, F=3, m_{F}=1\rangle$ \cite{PhysRevA.85.042310,bluvstein2021quantum} because it possesses a longer lifetime and mitigates the power requirements for the same Rabi frequency compared with the first resonance line in $5P$ state.
The simplified configuration of the atomic level is shown in Fig.~\ref{ple1}(c),
where we have introduced an uncoupled state $|d\rangle$ to denote the leakage level outside $|0\rangle$ and $|1\rangle$ for simplicity. Thus, the master equation of the system in Lindblad form reads
\begin{equation}\label{M1}
\frac{d\rho}{dt}=-i[H_{I},\rho]+\mathcal{L}_{p}[\rho]+\mathcal{L}_{r}[\rho],
\end{equation}
where
\begin{eqnarray}\label{H1}
H_{I}&=&\sum_{i=c,t}\frac{\Omega_{1}(t)}{2}|p\rangle_{i}\langle1|+\frac{\Omega_{2}}{2}|r\rangle_{i}\langle p|+{\rm H.c.}-\Delta|p\rangle_{i}\langle p|\nonumber\\&&+{\cal U}_{rr}|rr\rangle\langle rr|,
\end{eqnarray}
describes the coherent dynamics of the system,
and
\begin{equation}
\mathcal{L}_{p}[\rho]=\sum_{n=c,t}\sum_{i=0,1,d}L_{ip}^{(n)}\rho L_{ip}^{(n)^\dag}-\frac{1}{2}\{L_{ip}^{(n)^\dag}L_{ip}^{(n)},\rho\},
\end{equation}
\begin{equation}
\mathcal{L}_{r}[\rho]=\sum_{n=c,t}\sum_{j=0,1,d,p}L_{jr}^{(n)}\rho L_{jr}^{(n)^\dag}-\frac{1}{2}\{L_{jr}^{(n)^\dag}L_{jr}^{(n)},\rho\},
\end{equation}
picture the spontaneous emission from intermediate state $|p\rangle$ and Rydberg state $|r\rangle$, respectively with jump operator $L_{jr(ip)}^{(n)}$=$\sqrt{b_{jr(ip)}\gamma_{r(p)}}|j(i)\rangle_{n}\langle r(p)|$ and $b_{jr(ip)}$ denotes the branching ratio to the lower level $|j(i)\rangle$. At room temperature ($300~$K), the lifetime of state $|p\rangle$ and $|r\rangle$ are $\tau_{p}=1/\gamma_p=0.118~\mu$s and $\tau_{r}=1/\gamma_r=353~\mu$s, while the branching ratios are $b_{0(1)p}=1/8$, $b_{dp}=3/4$, $d_{1(0)r}=1/16$, $d_{dr}=3/8$, and $d_{pr}=1/2$. The term ${\cal U}_{rr}$ characterizes the vdW interaction of $-C_6/r^6$, and the second-order non-degenerate perturbation theory gives that the dispersion coefficient $C_{6}$ is about $-56.171~$THz$\cdot\mu$m$^{6}$ for Rydberg state $|100S_{1/2}\rangle$ \cite{SIBALIC2017319}. The reason why we choose $ns$ states instead of $nd$ states is that the interaction strength of $ns$ states is relatively isotropic, which is particularly important to maintain our system within the Rydberg blockade regime
when considering the thermal motion of atoms.

Now we discuss in detail the dynamic evolution of four input states for the truth table of a two-qubit $\textmd{CZ}_{\theta}$ gate, respectively. Since the ground state $|0\rangle$ is decoupled to the external driving fields, the input state $|00\rangle$  do not participate in the dynamics. The evolution form of the input two-atom states $|01\rangle$ and $|10\rangle$ are essentially the same as that of a single-atom state $|1\rangle$,
consequently
in what follows we only consider the asymmetric state $|01\rangle$ for the sake of convenience, and the Hamiltonian associated with it reads
\begin{equation}\label{h01}
H_{\textmd{eff}}^{(01)}=\frac{\Omega_{1}(t)}{2}|0p\rangle\langle01|+\frac{\Omega_{2}}{2}|0r\rangle\langle0p|+{\rm H.c.}-\Delta|0p\rangle\langle0p|,
\end{equation}
which has a dark instantaneous eigenstate $|\varphi(t)\rangle=\cos\vartheta|01\rangle-\sin\vartheta|0r\rangle$ with the mixing angle $\vartheta$=$\arctan[-\Omega_{1}(t)/\Omega_{2}]$. By properly modulating the shape of $\Omega_1(t)$ with time so that the amplitude of its initial time and final time are close to zero and satisfying the adiabatic approximation condition simultaneously,
\begin{equation}\label{ad1}
\bigg|\frac{\langle E_{0}^{01}(t)|\dot{E}_{\pm}^{01}(t)\rangle}{E_{\pm}^{01}(t)-E_{0}^{01}(t)}\bigg|
=\bigg|\frac{2\Omega_{2}(\dot{\Omega}_{1}(t)-\Omega_{1}(t))}{\Delta+\sqrt{\Delta^2+\Omega_{1}(t)^2+\Omega_{2}^2}}\bigg|\ll1,
\end{equation}
we can perform the cyclic evolution of state $|01\rangle$ without accumulating any geometric phase or dynamic phase.

For the case where the input state is $|11\rangle$, the analysis is somewhat complicated. In Fig.~\ref{ple1}(d), we give the transition path of relevant six symmetric states, where the population of state $|rr\rangle$ is suppressed due to the Rydberg blockade and the states $(|pr\rangle+|rp\rangle)/\sqrt{2}$ and $|pp\rangle$ are less populated for large detuning $2\Delta\gg\{\Omega_1(t)/2, \Omega_{2}/2\}$. Therefore, we can safely neglect these processes and the effective Hamiltonian can be written as
\begin{eqnarray}
H_{\textmd{eff}}^{(11)}&=&\frac{\sqrt{2}\Omega_{1}(t)}{2}|11\rangle\langle A|+\frac{\Omega_{2}}{2}|A\rangle\langle B|+{\rm H.c.}-\Delta|A\rangle\langle A|\nonumber\\&&+\frac{\Omega_{1}(t)^{2}}{4\Delta}|B\rangle\langle B|,
\end{eqnarray}
where $|A\rangle=(|1p\rangle+|p1\rangle)/\sqrt{2}$ and $|B\rangle=(|1r\rangle+|r1\rangle)/\sqrt{2}$. Compared with the coherent trapping type Hamiltonian of Eq.~(\ref{h01}), there is a time-dependent shift $\Omega_1(t)^2/4\Delta$ of state $|B\rangle$. Although this energy shift is very small within the parameter range we set, its existence will significantly modify the dynamics of the system, making the evolution completely different from the traditional coherent
trapping dynamics. The eigenvalues of $H_{\textmd{eff}}^{(11)}$ are the roots of the secular equation which appears as a cubic characteristic equation
\begin{equation}
E^{3}+aE^{2}+bE+c=0
\end{equation}
with $a=\Delta-\Omega_{1}(t)^2/4\Delta$, $b=-(3\Omega_{1}(t)^2+\Omega_{2}^2)/4$ and $c=\Omega_{1}^4/8\Delta$. The solutions to this cubic equation are
\begin{equation}
E_{0}^{11}(t)=\frac{2}{3}(-\frac{\Delta}{2}+\frac{\Omega_{1}(t)^{2}}{8\Delta}+\tilde{\Omega}\cos[\frac{\zeta}{3}]),
\end{equation}
\begin{equation}
E_{\pm}^{11}(t)=\frac{2}{3}(-\frac{\Delta}{2}+\frac{\Omega_{1}(t)^{2}}{8\Delta}+\tilde{\Omega}\cos[\frac{2\pi\mp\zeta}{3}]),
\end{equation}
with
\begin{equation}
\tilde{\Omega}=\frac{1}{2}[7\Omega_{1}(t)^{2}+3\Omega_{2}^{2}+4\Delta^{2}+\frac{\Omega_{1}(t)^{4}}{4\Delta^{2}}]^{1/2},
\end{equation}
\begin{eqnarray}
\zeta&=&2\pi-\arccos\{-[64\Delta^6-\Omega_{1}(t)^{6}+24\Delta^4(7\Omega_{1}(t)^2\nonumber\\&&+3\Omega_{2}^2)
+6\Delta^2(11\Omega_{1}(t)^4-3\Omega_{2}^2\Omega_{1}(t)^2)]/64\Delta^3\tilde{\Omega}^3\}.
\end{eqnarray}
The corresponding eigenvectors can be constructed as
\begin{widetext}
\begin{equation}
|E_{0}^{11}(t)\rangle=\cos\Theta|11\rangle+\sin\Phi\sin\Theta|A\rangle-\cos\Phi\sin\Theta|B\rangle,
\end{equation}
\begin{equation}
|E_{+}^{11}(t)\rangle=(\cos\Phi\cos\Theta\sin\phi+\sin\Phi\cos\phi)|B\rangle\nonumber-(\sin\Phi\cos\Theta\sin\phi-\cos\Phi\cos\phi)|A\rangle\nonumber+\sin\Theta\sin\phi|11\rangle,
\end{equation}
\begin{equation}
|E_{-}^{11}(t)\rangle=(\cos\Phi\cos\Theta\cos\phi-\sin\Phi\sin\phi)|B\rangle\nonumber-(\sin\Phi\cos\Theta\cos\phi+\cos\Phi\sin\phi)|A\rangle\nonumber+\sin\Theta\cos\phi|11\rangle,
\end{equation}
where
\begin{equation}
\Theta=\arctan\left[\frac{\Omega_{1}(t)[(E_{0}^{11}(t)-\Omega_{1}(t)^2/4\Delta)^2+\Omega_{2}^{2}/4]^{1/2}}{\sqrt{2}[(E_{0}^{11}(t)+\Delta)(E_{0}^{11}(t)-\Omega_{1}(t)^2/4\Delta)-\Omega_{2}^2/4]}\right],
\end{equation}
\end{widetext}
\begin{equation}
\Phi=\arctan\left[-\frac{2E_{0}^{11}(t)-\Omega_{1}(t)^2/2\Delta}{\Omega_{2}}\right],
\end{equation}
and it is not possible to find one expression for $\phi$  that is valid for all values of
the parameters \cite{1997Coherent,Shore1990}. Fortunately, this uncertainty does not affect our numerical simulation results below.
At $t=0$, $E_{0}^{11}(0)\rightarrow0$ and $|E_{0}^{11}(0)\rangle\approx\cos\Theta|11\rangle-\sin\Theta|B\rangle\approx|11\rangle$ because of $\Theta\approx\arctan[-\sqrt{2}\Omega_{1}(0)/\Omega_{2}]\approx0$.
\begin{figure}
\centering\scalebox{0.32}{\includegraphics{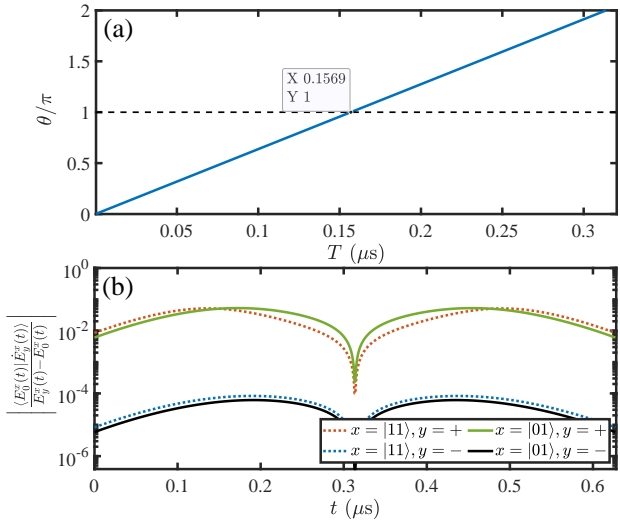}}
\caption{\label{Energy} (a) The relation between $\theta=\int_{0}^{4T}E_{0}^{11}(t)dt$ and $T$ under the parameters $\Omega_{0}/2\pi=160~$MHz, $\Omega_{2}/2\pi=200~$MHz and $\Delta/2\pi=1000~$MHz, where $\int_{0}^{4T}E_{0}(t)dt=\pi$ at $T\approx0.157~\mu$s. (b) The variation in adiabatic conditions of the system under the same parameters with $T=0.157~\mu$s.}
\end{figure}
Therefore under the adiabatic evolution condition of this case
\begin{equation}\label{ad2}
\bigg|\frac{\langle E_{0}^{11}(t)|\dot{E}_{\pm}^{11}(t)\rangle}{E_{\pm}^{11}(t)-E_{0}^{11}(t)}\bigg|\ll1,
\end{equation}
the state $|11\rangle$ evolves along the eigenstate $|E_{0}^{11}(t)\rangle$ from the beginning to the end as
\begin{equation}
|\Psi(t)\rangle=e^{-i\int_{0}^{t}E_{0}^{11}(t')dt'}|E_{0}^{11}(t)\rangle,
\end{equation}
from which a dynamical phase $-\int_{0}^{T_g}E_{0}^{11}(t')dt'$ is acquired after state $|11\rangle$ undergoing a cyclic evolution over the gate operation time $T_g$.
\begin{figure*}
\centering\scalebox{0.33}{\includegraphics{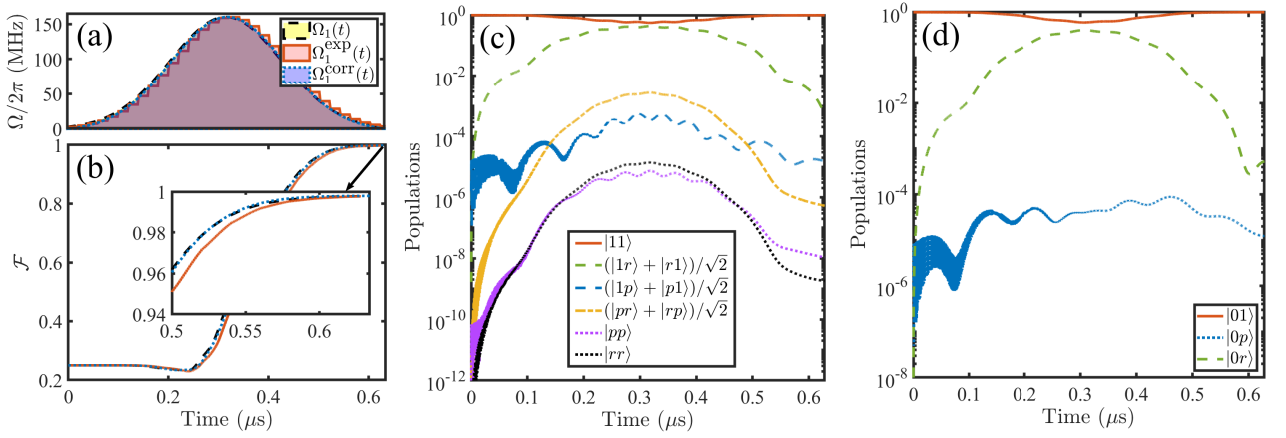}}
\caption{\label{Evolution}The realization of the CZ gate governed by the master equation (\ref{M1}). (a) The time dependence of Rabi frequency of application, where $\Omega_{1}(t)$, $\Omega_{1}^{\textmd{exp}}(t)$ and $\Omega_{1}^{\textmd{corr}}(t)$ correspond to standard Gaussian pulse, experimental Gaussian pulse and corrected Gaussian pulse, respectively. (b) The fidelity of the CZ gate corresponding to the above three pulses. (c) Populations of states $|11\rangle$, $|pp\rangle$, $(|1r\rangle+|r1\rangle)/\sqrt{2}$, $(|1p\rangle+|p1\rangle)/\sqrt{2}$, $(|pr\rangle+|rp\rangle)/\sqrt{2}$, $|rr\rangle$ for the initial state $|11\rangle$ with $\Omega_{1}(t)$. (d) Populations of the $|01\rangle$, $|0p\rangle$ and $|0r\rangle$ states for initial state $|01\rangle$ with $\Omega_{1}(t)$. The parameters are taken as $\Omega_{0}/2\pi=160~$MHz, $\Omega_{2}/2\pi=200~$MHz, $\Delta/2\pi=1000~$MHz, ${\cal U}_{rr}/2\pi=2~$GHz, $T=0.157~\mu$s, $\tau_{r}=353~\mu$s and $\tau_{p}=0.118~\mu$s.}
\end{figure*}
Remarkably, there are many options for time-dependent calibrated pulses that meet the condition of our scheme.
For the convenience of experimental implementation, we here take the time-dependent Rabi frequency $\Omega_{1}(t)$ as a Gaussian pulse in the form of
\begin{equation}
\Omega_{1}(t)=\Omega_{0}e^{-\frac{(t-2T)^{2}}{T^{2}}},
\end{equation}
 where
$\Omega_{0}$ and $T$ are the maximum amplitude and width of the
Gaussian pulse, respectively. On the basis of this, the evolution time of the system should be set as $T_g=4T$ since the pulse $\Omega_{1}(t)$ peaks at $t=2T$. In order to realize the two-qubit controlled arbitrary-phase $\textmd{CZ}_{\theta}$ gate, we need
\begin{equation}\label{T1}
\int_{0}^{4T}E_{0}^{11}(t')dt'=\theta,
\end{equation}
where the phase factor $\theta$ can be adjusted arbitrarily in the range of 0 to $\pi$.
From the above analyses, we know that only $|11\rangle$ will accumulate an effective dynamic phase through the non-zero eigenenergy. Thus, to determine the adjustable evolution time $4T$, we have to get an integral expression of Eq.~(\ref{T1}). However, due to the complicated form of $E_{0}^{11}(t)$, the analytic form of the integral is difficult to calculate, so we instead resort to the numerical integration method by scanning the results with different $T$ and try to find the point where the integral is $\theta$, as shown in Fig.~\ref{Energy}(a).

Considering $\theta=\pi$ as an example, starting from initial state $|\Psi(0)\rangle=(|00\rangle+|01\rangle+|10\rangle+|11\rangle)/2$, the fidelity  $\mathcal{F}$ of the standard CZ gate is defined by the population of the target state $|\Psi_{t}\rangle=(|00\rangle+|01\rangle+|10\rangle-|11\rangle)/2$. It should be noted that the definition of gate fidelity used here is essentially the same as the definition of Bell-state fidelity used in the previous literature \cite{PhysRevLett.123.170503,PhysRevA.101.062309}. To achieve the strong Rydberg blockade, we choose ${\cal U}_{rr}/2\pi=2~$GHz corresponding to an interatomic spacing $r\simeq5.5~\mu$m. Moreover, by fixing the parameters $\Omega_{2}/2\pi=200~$MHz and $\Delta/2\pi=1000~$MHz, the relationship among the fidelity of the CZ gate, the evolution time $4T$, and the parameter $\Omega_{0}$ is shown in Table.~\ref{parameter} governed by Eq.~(\ref{M1}).
Theoretically speaking, for a smaller $\Omega_{0}$, $\mathcal{F}_{t}$ can reach over $0.9999$ without considering the spontaneous emissions. However, this condition results in a long evolution time  that may deepen the influences of spontaneous emissions and dephasing for a realistic situation. For the above reasons, unless otherwise specified, we select $\Omega_{0}/2\pi=160~$MHz in the following analysis to implement a relatively fast and high-fidelity logic gate.
\begin{table}\label{Tp}
\centering
\caption{\label{parameter}The relationship among the fidelity of the CZ gate, the evolution time $4T$, and the maximal pulse amplitude $\Omega_{0}$. The other parameters are taken as $\Omega_{2}/2\pi=200~$MHz, $\Delta/2\pi=1000~$MHz, ${\cal U}_{rr}/2\pi=2~$GHz.}
\setlength{\tabcolsep}{2mm}
\begin{tabular}{cccccc}
\hline\hline
$\Omega_{0}/2\pi~$(MHz)&80&100&120&140&160\\
$4T~(\mu$s)&5.994&2.7956&1.5352&0.9428&0.628\\
\hline
$\mathcal{F}_{t}(\gamma=0)$&0.9999&0.9999&0.9997&0.9993&0.9990 \\
$\mathcal{F}_{t}(\gamma\neq0)$&0.9982&0.9984&0.9984&0.9980&0.9978 \\
\hline\hline
\end{tabular}
\end{table}
According to the relevant levels of $^{87}$Rb, the $|1\rangle\leftrightarrow|p\rangle$ transition is driven by a $420~$nm beam with typical beam power $P_{0}=78.5~\mu$W and waist of $\omega_{x|y,0}=4.2~\mu$m which gives a Rabi frequency $\Omega_{0}/2\pi=160~$MHz. By tuning the $1011~$nm beam with typical beam power $P_{2}=290.5~$mW and waist of $\omega_{x|y,2}=3.9~\mu$m, the Rabi frequency of $\Omega_{2}/2\pi=200~$MHz can be realized to coupe the transition of $|p\rangle\leftrightarrow|r\rangle$. As shown in Fig.~\ref{Energy}(a), after scanning the numerical integration results, we have $T=0.157~\mu$s under such parameters, and
Fig.~\ref{Energy}(b) shows the variation in adiabatic conditions of the system given by Eqs.~(\ref{ad1}) and (\ref{ad2}) versus time.  In the evolution process, these values are always far less than 1, which ensures a nearly perfect coherent population transfer process.

In Fig.~\ref{Evolution}(a) and \ref{Evolution}(b), we first discuss the system dynamics driven by the Gaussian pulse $\Omega_{1}(t)$. Under the domination of the master equation Eq.~(\ref{M1}), we can obtain the CZ gate with a fidelity of 0.9978  (dashed line) within 1~$\mu$s operation time. Since the Gaussian function may introduce an extra disadvantage due to the non-vanishing tail, we then make a correction on the standard pulse by employing $\Omega_{1}^{\textmd{corr}}(t)=\Omega_{0}[e^{-(t-2T)^{2}/T^{2}}-a]/(1-a)$, where $T=0.1585~\mu$s and $a$ is set to give an exact zero amplitude at the start and the end of the Gaussian pulse, and the corresponding gate fidelity is 0.9979  (dotted line), which means the error caused by the non-vanishing tail of the Gaussian pulse has little effects on the fidelity of our scheme. In fact, the temporal pulse can be adopted with no need for a strict zero amplitude at the start and the end, but small enough to ensure that the adiabatic condition is established.
To be more realistic, we also numerically simulated the system dynamics under the experimentally available pulse $\Omega_{1}^{\textmd{exp}}(t)$ composed of about 31 cylindrical pulses with a duration $0.02~\mu$s and amplitudes $\Omega_{1}(0.02n)$ ($n=0,1...,30$). In this case, the gate fidelity can still reach 0.9977 (solid line). Therefore, the above results show that the Gaussian pulse form is consistent with experimental and theoretical predictions. Figs.~\ref{Evolution}(c) and \ref{Evolution}(d) depict the dynamics of each input state with $\Omega_{1}(t)$ in detail, and confirm that  in the process of realizing the CZ gate, the symmetric states $|rr\rangle$, $(|pr\rangle+|rp\rangle)/\sqrt{2}$, and $|pp\rangle$ are well suppressed.


For a controlled arbitrary-phase gate, we still use the population of the target state $|\Psi_{t}'\rangle=(|00\rangle+|01\rangle+|10\rangle+e^{-i\theta}|11\rangle)/2$ starting from $|\Psi(0)\rangle$ as the definition of the gate fidelity. It is noteworthy that there is no necessary to discuss the situation for an  extremely small phase $\theta$, since the $\textmd{CZ}_{\theta}$ gate gets very closed to the unit operator in this case, i.e. $|\textmd{Tr}[U_{\textmd{I}}^{\dag}U_{\textmd{CZ}_{\theta}}]|^2/16$= $1-3\theta^2/16+\mathcal{O}[\theta^4]$.
Considering the error of experimental operation and atomic spontaneous emission, it is better to ``realize" a small-angle controlled-phase gate without any operation.
When the wanted phase exceeds $0.08\pi$, the quantity $|\textmd{Tr}[U_{\textmd{I}}^{\dag}U_{\textmd{CZ}_{\theta}}]|^2/16$ drops below 0.99,
and this is the scope of the beginning of the phase we are interested in discussing. In Fig.~\ref{phase}, we take into account the trade-off between the Rabi frequency and the pulse duration, and plot the fidelities of different $\textmd{CZ}_{\theta}$ gates under multiple sets of parameters. The inset of Fig.~\ref{phase} retains the selectable pulse and the corresponding operation time for different phases. In Table.~\ref{Controlled-phase}, we also list the optimal parameters of the Gaussian pulse  corresponding to $\theta\in[0.08\pi,\pi]$ for reference, and a high-fidelity continuous controlled-phase gate set with operation time less than $1~\mu$s can be obtained under these parameters.

{\begin{figure}
\centering\scalebox{0.32}{\includegraphics{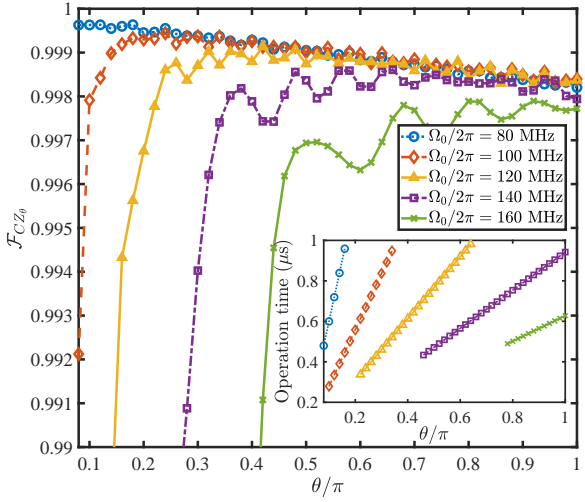}}
\caption{\label{phase}The fidelity of CZ$_\theta$ gate with different parameters $\Omega_{0}$. The inset shows the corresponding evolution time under above situation corresponding to the selectable pulses. The other parameters are $\Omega_{2}/2\pi=200~$MHz, $\Delta/2\pi=1000~$MHz, and ${\cal U}_{rr}/2\pi=2~$GHz.}
\end{figure}
\begin{table}
\centering
\caption{\label{Controlled-phase}A reference for the parameter choices for CZ$_{\theta}$ gate with fidelity over 0.997 and the evolution time less than $1~\mu$s. The other parameters are taken as $\Omega_{2}/2\pi=200~$MHz, $\Delta/2\pi=1000~$MHz, and ${\cal U}_{rr}/2\pi=2~$GHz.}
\setlength{\tabcolsep}{4.5mm}
\begin{tabular}{ccc}
\hline\hline
$\theta$~(rad)&$\Omega_{0}/2\pi~$(MHz)&$T(\theta)~(\mu$s) \\
\hline
$0.08\pi-0.16\pi$&80&$T=\theta/2.0965$ \\
$0.1\pi-0.34\pi$&100&$T=\theta/4.5064$ \\
$0.22\pi-0.64\pi$&120&$T=\theta/8.2$\\
$0.34\pi-\pi$&140&$T=\theta/13.337$\\
$0.64\pi-\pi$&160&$T=\theta/20.029$ \\
\hline\hline
\end{tabular}
\end{table}}
\section{Applications to Max-cut problem and Fourier transform}\label{secnew}
\begin{figure}
\centering\scalebox{0.23}{\includegraphics{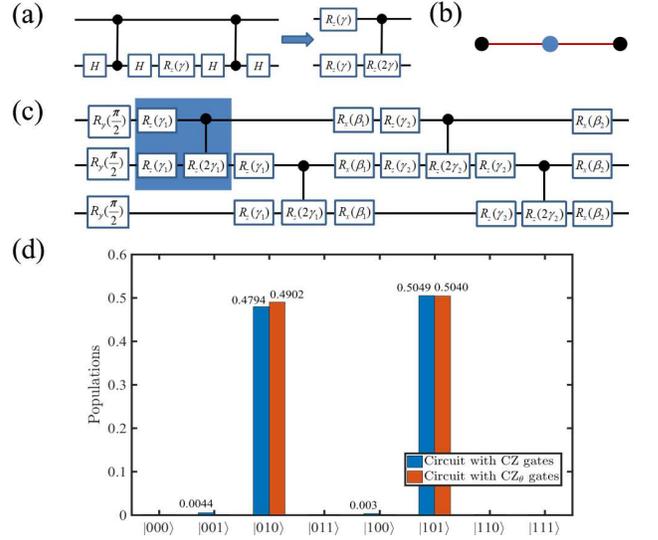}}
\caption{\label{Max-Cut}Quantum circuits and numerical simulation results for the Max-Cut problem of a three-vertex line graph. (a) The simplification of the circuit unit with the application of the CZ$_\theta$ gates. (b) The three-vertex line graph with the centre vertex connected to the two sides vertices. (c) The complete quantum circuit used to solve the Max-Cut problem above, which is constructed by CZ$_\theta$ gates. (d) The numerical simulation results under the circuit constructed by CZ and CZ$_\theta$ gates, respectively. The input state is $|000\rangle$ and the parameter for the CZ$_\theta$ gates are chosen according to Table.~\ref{Controlled-phase}, where $\Omega_{0}/2\pi=140~$MHz.}
\end{figure}

The Max-Cut problem attempts to partition the vertices of a graph into two sets so that the maximum number of edges can be cut. As a typical NP-hard problem, it can be solved by mapping the problem of finding the ground state of a cost Hamiltonian $H_c=1/2\sum_{\alpha=1}^{m}(1-\sigma^z_{\alpha_{1}}\sigma^z_{\alpha_{2}})$, in which $\sigma^z$ is the
Pauli $Z$ operator and $\alpha_{1,2}$ are qubit indices representing the vertices of the edge $\alpha$. The standard  method of quantum circuit for QAOA to find the ground state of $H_c$ is measuring the probability distribution of the final state $|\gamma\beta\rangle=U_{m}(\beta_{p})U_{c}(\gamma_{p})\cdots U_{m}(\beta_{1})U_{c}(\gamma_{1})|s\rangle$  on the computational basis \cite{PhysRevApplied.14.034010,Graham2022}, where $|s\rangle=[1/\sqrt{2} (|0\rangle+|1\rangle)]^{\otimes N}$ is the initial state, and $U_{c}=e^{i\gamma H_{c}}$ and $U_{m}=e^{i\beta H_{m}}$ ($H_{m}=-1/2\sum_{\alpha=1}^{m}\sigma_{\alpha}^{x}$) represent the cost function and state mixing, respectively. The optimal settings for all $\gamma_{i}$ and $\beta_{i}$ can be found via the classical optimizer.
In the limit of $p\rightarrow\infty$, the above result can be regarded as a Trotterized version of adiabatic evolution of the initial state to the ground state of $H_{c}$.

As an illustration, the equivalent quantum circuit is displayed in Fig.~\ref{Max-Cut}(c), which depicts the Max-Cut problem on a three-vertex line in Fig.~\ref{Max-Cut}(b), where the application of the CZ$_\theta$ gate reduces the two-qubit terms in the quantum circuit as shown in Fig.~\ref{Max-Cut}(a) by a factor of two when compared to the Refs.~\cite{PhysRevApplied.14.034010,Graham2022}. There are two degenerate Max-Cut solutions for the three-vertex line graph ($|101\rangle$ and $|010\rangle$) where the center vertex is connected to the two side vertices and the two cuts result from the center and side vertices belonging to distinct sets.
The outcomes of this graph under the quantum circuits built by CZ gate and CZ$_\theta$ gate, respectively, are described in Fig.~\ref{Max-Cut}(d). The optimized parameters are about $\gamma\simeq\{0.338\pi, 0.559\pi\}$ and $\beta\simeq\{0.669\pi, -0.228\pi\}$. We assume that all single-qubit gates are perfect, but the equivalent two-qubit gates are formed by evolution contains the error cased by the spontaneous emission under Eq.~(\ref{M1}), in order to compare the difference caused by the two-qubit gates in the above two quantum circuits.
When layer $p=2$ is utilized, both circuits yield results that are reasonably accurate, with fidelities of $\tilde{F}_{CZ}=0.9840$ and $\tilde{F}_{CZ_\theta}=0.9940$, respectively.
The overall evolution times for the circuits without considering single-qubit gates are $t_{CZ}\simeq5.024~\mu$s and $t_{CZ_{\theta}}\simeq3.383~\mu$s, respectively. In more complex systems, additional layers will be required for the right solution, and two-qubit gate errors will also build up. Therefore, it is possible that the application of CZ$_\theta$ gate can be used to address more complex problems, improve accuracy, and do so while utilizing half as many two-qubit gates and necessitating a shorter running time \cite{PRXQuantum.1.020304}.

\begin{figure}
\centering\scalebox{0.23}{\includegraphics{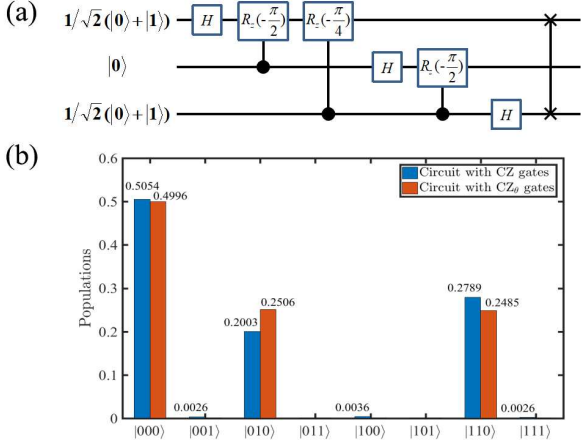}}
\caption{\label{Fourier}(a) The Quantum circuit for the Fourier transform constructed by CZ$_\theta$ gates. (b) The numerical simulation results of the quantum circuit with CZ and CZ$_\theta$ gates, respectively. The input state is $1/\sqrt{2}(|0\rangle+|1\rangle)\otimes|0\rangle\otimes1/\sqrt{2}(|0\rangle+|1\rangle)$. The parameters $\Omega_{0}/2\pi=\{120,100\}~$MHz are chosen to obtain the CZ$_{\theta}$ gates with $\theta=\{-\pi/2,-\pi/4\}$.}
\end{figure}

The continuous control phase gate is necessary for the standard Fourier transform problem in quantum computing. Fig.~\ref{Fourier}(a) depicts the complete quantum circuit and Fig.~\ref{Fourier}(b) describes the results of the Fourier transform with input state $1/\sqrt{2}(|0\rangle+|1\rangle)\otimes|0\rangle\otimes1/\sqrt{2}(|0\rangle+|1\rangle)$ under the circuit of CZ gate and CZ$_\theta$ gate, respectively. It is important to note that we modify  the detuning to be $\Delta= -2\pi\times1000$~MHz in Eq.~(\ref{M1}) in order to produce the required negative phase.
Compared with the ideal result $1/\sqrt{2}[|000\rangle+0.5(1+i)|010\rangle+0.5(1-i)|110\rangle]$, the fidelities of the two quantum circuits are $\tilde{F}_{CZ}=0.9808$ and $\tilde{F}_{CZ_\theta}=0.9987$, demonstrating that the CZ$_\theta$ gates-applied circuit is more accurate. The corresponding evolution time of the circuits without taking into account the swap gate and the single-qubit gates are $t_{CZ}\simeq3.768~\mu$s and $t_{CZ_{\theta}}\simeq2.23~\mu$s, respectively.

Based on the analysis presented above, it is clear that creating a continuous control phase gate successfully can simplify quantum circuits and increase the precision with which complex quantum computing problems are resolved.

\section{Discussion of the experimental feasibility and technical imperfections}\label{sec3}
In experiments, the magneto-optical trap (MOT) technology based on the Doppler cooling mechanism is the most commonly used laser cooling and trapping method, it provides a platform for many fundamental research and applications using cold atomic systems \cite{Barry2014,science.aan5614}. However, since MOT cannot store the quantum state for a long time, it is necessary to introduce additional capture methods without affecting the control of quantum states. Another more important reason is that the atomic cooling and trapping scales of MOT vary from several hundred microns to several millimeters, which is much larger than the Rydberg blocking radius. Therefore, in the experiment, neutral atom are loaded into the  far-off-resonance optical
traps (FORTs) \cite{PhysRevA.72.022347,PhysRevA.85.042310} or  optical tweezers \cite{PhysRevA.78.033425,PhysRevX.2.041014,Brooks_2021} via MOT to achieve further capture. The trapping potential of a far-off-resonant optical tweezer with linearly polarized light can be described by \cite{GRIMM200095,WALKER201281}
\begin{equation}
U_{F}(\mathbf{r})=\frac{\pi c^{2}\Gamma}{2\omega_{0}^{3}}(\frac{2}{\Delta_{3/2}}+\frac{1}{\Delta_{1/2}})I(\mathbf{r}),
\end{equation}
where $\omega_{0}$ and $\Gamma$ are the frequency and decay rate of $5S_{1/2}-5P_{3/2}$ transition and $\Delta_{3/2(1/2)}$ is the laser detuning from the $5P_{3/2(1/2)}$. The trap depth can be calculated with the presence of peak trapping intensity $I(0)=2P_{f}/\pi\omega_{f}^2$, where $P_{f}$ and $\omega_{f}$ are respectively the power and the waist of the tweezer beam.
By applying polarization-gradient cooling and adiabatic, the experimental apparatus in Ref.~\cite{PhysRevA.105.042430} can cool the atomic temperature to $5.2~\mu$K in a $50~\mu$K ($U_{F}/k_{B}$) trap. In order to make our scheme consistent with the data provided by this trap, the parameters of the laser beams are set as wavelength $\lambda_{f}=830~$nm, the typical beam power $P_{f}=174~\mu$W, and the waist ($1/e^{2}$ intensity radius) $\omega_{f}=1.2~\mu$m.

As studied in Refs.~\cite{PhysRevA.101.043421,PhysRevA.85.042310,PhysRevA.97.053803,PhysRevLett.123.230501}, we conduct numerical analysis on the technical imperfections of realizing the CZ gate from four aspects: (i) Doppler shift and fluctuation of Rydberg-Rydberg interaction strengths, (ii) inhomogeneous Rabi frequency, (iii) fluctuation and noise of external fields, and (vi) finite detection errors. The detailed analyses are listed below in subsections. To be more credible, all results are averaged over 100 realizations referring to the fluctuations of the above parameters.

\subsection{Doppler shifts and fluctuations of the Rydberg-Rydberg interaction strength}
Due to the limitation of the existing cooling mechanism, the temperature of the atom cannot reach absolute zero. Therefore, the atom has a certain speed  leading to the Doppler effect, and the laser frequency detuning felt by the atom will be shifted from the desired $\Delta$. Moreover, atoms affected by non-zero temperature will cause vibrations near the ideal position. Combining these two reasons, the actual distance ${\it l(t)}$ between
the pair of atoms varies with time, resulting in fluctuations in the Rydberg-Rydberg interaction.
The ideal position of the control and target atoms are denoted as $\mathbf{R}_{c}=(0,0,0)$ and $\mathbf{R}_{t}=(r,0,0)$, respectively. The Hamiltonian includes atomic motion and fluctuation of vdW interaction is
\begin{eqnarray}
H_{v}&=&\sum_{i=c,t}\frac{\Omega_{1}(t)}{2}e^{i\mathbf{k}_1\cdot\mathbf{R}_i(t)}|p\rangle_{i}\langle1|+\frac{\Omega_{2}}{2}e^{i\mathbf{k}_{2}\cdot\mathbf{ R}_{i}(t)}|r\rangle_{i}\langle p|\nonumber\\&&+{\rm H.c.}-\Delta|p\rangle_{i}\langle p|+{\cal U}_{rr}[\it{l(t)}]|rr\rangle\langle rr|,
\end{eqnarray}
where $\mathbf{R}_{i}(\tau)=\mathbf{R}_{i}+\delta\mathbf{R}_{i}+\mathbf{v}_{i}t$ and ${\it l(t)}=|\mathbf{R}_{c}(t)-\mathbf{R}_{t}(t)|$. The randomly generated three-dimensional position $\delta\mathbf{R}_{i}$ and velocity vector $\mathbf{v}_{i}$ obey the Maxwell-Boltzmann distribution \cite{PhysRevApplied.13.024008}.
The time-averaged variances of atomic position and momentum are shown as \cite{PhysRevA.72.022347}
\begin{equation}
\langle x^2\rangle=\langle y^{2}\rangle=\frac{\omega_{f}^{2}}{4}\frac{T_{a}}{|U_{F}|},~~\langle z^{2}\rangle=\frac{\pi^{2}\omega_{f}^{4}}{2\lambda_{f}^{2}}\frac{T_{a}}{|U_{F}|},
\end{equation}
\begin{equation}
\langle v_{x}^{2}\rangle=\langle v_{y}^{2}\rangle=\langle v_{z}^{2}\rangle=\frac{T_{a}}{m},
\end{equation}
where $T_{a}$ is the measured temperature of the trapped atoms. In our setup [Fig.~\ref{ple1}(a)], the two excitation lasers with vectors  $\mathbf{k}_{1}$ and $\mathbf{k}_{2}$ are counter-propagating along $z$-axis. The Hamiltonian can be rewritten as
\begin{eqnarray}\label{Hv}
H_{v}&=&\sum_{i=c,t}\frac{\Omega_{1}(t)}{2}e^{ik_{1}^{z}Z_i(t)}|p\rangle_{i}\langle1|+\frac{\Omega_{2}}{2}e^{-ik_{2}^{z}Z_i(t)}|r\rangle_{i}\langle p|\nonumber\\&&+{\rm H.c.}-\Delta|p\rangle_{i}\langle p|+{\cal U}_{rr}[{\it l}(t)]|rr\rangle\langle rr|,
\end{eqnarray}
where $Z_{i}=z_{i}+\delta z_{i}+v_{zi}t$. The corresponding wave vectors are $k_{1}^z/2\pi\simeq2.381\times10^6/$m and $k_{2}^z/2\pi\simeq0.989\times10^6/$m. The vdW interaction becomes ${\cal U}_{rr}[{\it l}(t)]/2\pi=-C_{6}/{\it l}(t)^{6}$. Assuming the position distribution and velocity vector of two atoms are both Gaussian with variance of
\begin{equation}
\sigma_{x}^{2}=k_{B}\langle x^{2}\rangle,~~\sigma_{y}^{2}=k_{B}\langle y^{2}\rangle,~~\sigma_{z}^{2}=k_{B}\langle z^{2}\rangle,
\end{equation}
\begin{equation}
\sigma_{v_{x}}^{2}=k_{B}\langle v_{x}^{2}\rangle,~~\sigma_{v_{y}}^{2}=k_{B}\langle v_{y}^{2}\rangle,~~\sigma_{v_{z}}^{2}=k_{B}\langle v_{z}^{2}\rangle,
\end{equation}
with $k_{B}$ the Boltzmann constant.
The random number subject to Gaussian distribution can be generated with two uniformly distributed random numbers $\xi$ in the interval $[0,1]$, denoted as $\sigma_{i}\sqrt{-2\ln\xi_{1}}\cos[2\pi\xi_{2}]$.
\begin{figure}
\centering\scalebox{0.35}{\includegraphics{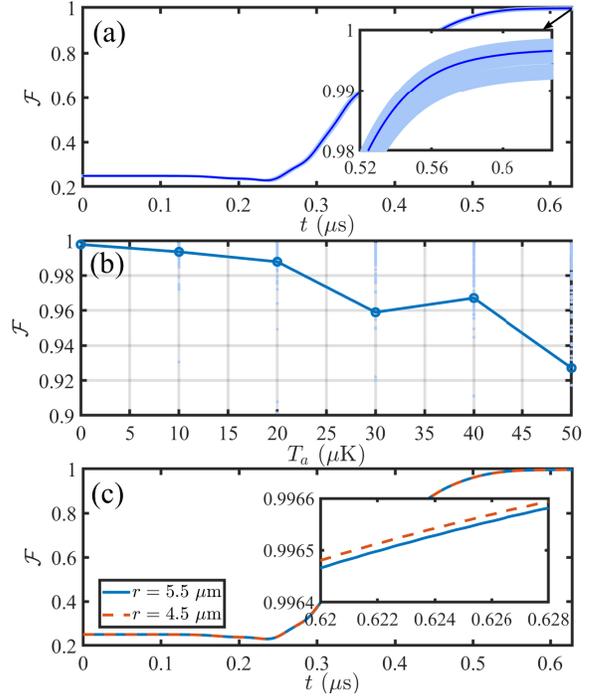}}
\caption{\label{Doppler}(a) The system dynamics considering the Doppler effect and the fluctuation of vdW interaction at the finite temperature $T_{a}=5.2~\mu$K governed by the master equation (\ref{M1}) with Hamiltonian (\ref{Hv}). (b) The gate fidelity at different temperature $T_{a}$. (c) The average evolution results with initial separation $r=5.5~\mu$m (solid line) and $r=4.5~\mu$m (dashed line), respectively. Note that the light blue region show the results of a hundred stochastic simulations and the solid line in dark blue corresponds to the average results. The parameters are taken as same as Fig.~\ref{Evolution}.}
\end{figure}

Fig.~\ref{Doppler}(a) portrays the fidelity of CZ gate governed by the master equation with Hamiltonian ($\ref{Hv}$) under $T_{a}=5.2~\mu$K. The light blue parts represent the results of a hundred times stochastic simulations and the solid line in dark blue corresponds to the average result. The average gate error is about $0.00118$. Fig.~\ref{Doppler}(b) shows the fidelity of present gate protocol versus atomic temperatures $T_{a}$, indicating that lower cooling temperatures facilitate the generation of gates. The reason is that ${\cal U}_{rr}$ is
related to the atomic separation ${\it l}(t)$. With the increase of $T_{a}$, the range of atomic motion expands, which cannot guarantee the strong Rydberg blockade and lead to a greater error. According to our setup, the gate fidelity can hold above $0.98$ with $T_{a}<20~\mu$K.
In Fig.~\ref{Doppler}(c), we analyze the evolution results of $T_{a}=5.2~\mu$K on average at $r=5.5~\mu$m and $r=4.5~\mu$m, respectively,  illustrating that the error can be further reduced by reducing the initial distance between atoms.
In addition, under the influence of atomic temperature, the vibration of atoms in the direction of tweezers beam is more intense. Therefore, we arrange atoms perpendicular to the tweezers beam to reduce the effect of atomic vibration.

\subsection{Inhomogeneous Rabi frequency}
In the above section, we have discussed the influence of Doppler shifts and fluctuations of the vdW interaction caused by atomic vibrations at finite temperature. But subject to the beam waists of lasers, the vibration will also make the atoms deviate from the laser center, resulting in changes in the actual optical intensity felt by the atoms. The reduction of the Rabi frequency has been found when atoms are prepared at a distance from the addressed site. In Ref.~\cite{Christandl2016}, the spatial dependence of Rabi frequency has been numerically studied, from which we have the position-dependent Rabi frequencies  \cite{PhysRevA.85.042310}
\begin{equation}
\Omega_{1}(t,\mathbf{\tilde{R}_{i}})=\Omega_{1}(t,0)\frac{e^{-[\frac{x^{2}}{\omega_{x,0}^{2}(1+z^{2}/L_{x,0}^{2})}+\frac{y^{2}}{\omega_{y,1}^{2}(1+z^{2}/L_{y,0}^{2})}]}}{[(1+z^{2}/L^{2}_{x,0})(1+z^{2}/L^{2}_{y,0})]^{1/4}},
\end{equation}
\begin{eqnarray}
\Omega_{2}(\mathbf{\tilde{R}_{i}})&=&\Omega_{2}(0)\frac{e^{-[\frac{x^{2}}{\omega_{x,2}^{2}(1+z^{2}/L_{x,2}^{2})}+\frac{y^{2}}{\omega_{y,2}^{2}(1+z^{2}/L_{y,2}^{2})}]}}{[(1+z^{2}/L^{2}_{x,2})(1+z^{2}/L^{2}_{y,2})]^{1/4}},
\end{eqnarray}
where $\Omega_{1}(t,0)$ and $\Omega_{2}(0)$ are the Rabi frequencies at trap center, $L_{x|y,i}=\pi\omega_{x|y,i}^2/\lambda_{i}$ is the Rayleigh length. The trap position of atom $i$ is $\mathbf{\tilde{R}}_{i}=\mathbf{\tilde{R}}_{i}+\delta\mathbf{R}_{i}$, where $\mathbf{\tilde{R}}_{i}$ is the ideal position denoting the laser alignment. Because two atoms are driven independently, the definition of $\mathbf{\tilde{R}}_{i}$ equals $(0,0,0)$ independent of the relative position of atoms.
After 100 repeated numerical simulations, it is found that when the atomic temperature $T_{a}=5.2~\mu$K, the influence of Rabi frequency inhomogeneity on the system dynamics is only 0.00044.


\subsection{Fluctuation and noise of external fields}

Usually, multiple fields need to be applied in the experiments of neutral-atom systems, such as the laser field used to drive the atom and the magnetic field used to lift the degeneracy of the Zeeman sublevels. The gate errors introduced by the fluctuation and noise of these external fields will be discussed in this section.

(i) The fluctuation of the Rabi frequency. The intensity fluctuation of laser fields will introduce a time-dependent fluctuation $\delta\Omega_{i}(t)$ on the driving Rabi frequency, which is assumed to follow the normal distribution functions with the standard deviations $\sigma_{\Omega_{1,2}}\approx0.05\Omega_{0,2}$. Then the system Hamiltonian reads
\begin{eqnarray}\label{HO}
H_{\Omega}&=&\sum_{i=c,t}\frac{1}{2}[\Omega_{1}(t)+\delta \Omega_{1}(t)]|p\rangle_{i}\langle1|+\frac{1}{2}[\Omega_{2}+\delta\Omega_{2}(t)]|r\rangle_{i}\langle p|\nonumber\\&&+{\rm H.c.}-\Delta|p\rangle_{i}\langle p|+{\cal U}_{rr}|rr\rangle\langle rr|.
\end{eqnarray}
As shown in Table.~\ref{err}, the fluctuation of Rabi frequency has little effect on the system which is about 0.0001.

\begin{figure}
\centering\scalebox{0.32}{\includegraphics{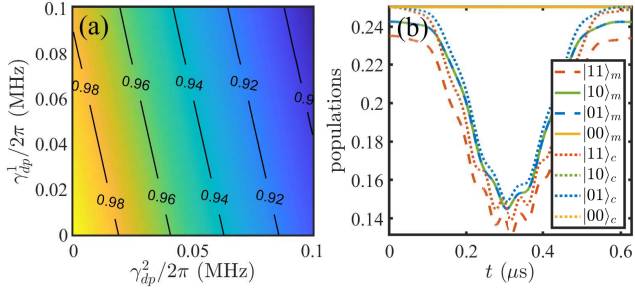}}
\caption{\label{Error}(a) The gate fidelities with laser phase noise governed by Eq.~(\ref{M1}) plus (\ref{dephase}). (b) The measured and corrected populations of states $|11\rangle$, $|10\rangle$ $|01\rangle$ and $|00\rangle$ with initial state $|\Psi(0)\rangle=1/2(|00\rangle+|01\rangle+|10\rangle+|11\rangle)$ and finite detection errors $(\epsilon,\epsilon')=(0.03,0.0047)$.}
\end{figure}
\begin{table*}
\centering
\caption{\label{err}Fidelity errors of the CZ gate corresponding to the system constructed by $^{87}$Rb atoms trapped in two optical tweezers which show in Ref.~\cite{PhysRevA.105.042430} relative to the ideal fidelity $0.9990$. The tweezers are generated by the tightly focused $830~$nm laser, with beam waist at focal plane $1.2(1)~\mu$m with trap depth $50~\mu$K and the temperature of single-atom is about $T_{a}=5.2~\mu$K. The lower section gives the average results after a hundred times numerical simulation.}
\setlength{\tabcolsep}{4.5mm}
\begin{tabular}{ccc}
\hline\hline
Quantity&Relative error budget&Fidelity estimate\\
\hline
Spontaneous emission &0.00124& \\
Doppler effects and the fluctuation of Rydberg-Rydberg interaction strengths&0.00118& \\
The inhomogeneous Rabi frequency&0.00044\\
The fluctuation of Rabi frequency&0.0001& \\
Laser noises ($\gamma_{dp}/2\pi=10~$kHz)&0.01151& \\
Fluctuation of detuning&-0.000006&$\mathcal{F}_{Cz}\simeq0.984616$ \\
\hline
Detection errors&$0.01\sim0.03$&$\mathcal{F}_{Cz}\simeq0.974616\sim0.954616$\\
\hline\hline
\end{tabular}
\end{table*}

(ii) The phase noise of laser fields. The laser phase noise can be  written as $\Omega_{i}(t)=\Omega_{i}\exp(i\varphi_{i}(t))$, where $\varphi_{i}(t)$ presents as a random process related to the power spectral density $S_{\varphi}(f)$ with phase-modulated Fourier frequency $f$. Because $S_{\varphi}(f)$ depends on the test results of specific experiments, the laser phase noise is difficult to quantify directly \cite{PhysRevA.97.053803,PhysRevA.99.043404}. Fortunately, the average result of the laser phase noise will lead to dephasing of Rabi oscillations \cite{PhysRevA.101.043421,Madjarov2020}, and it can be described as
\begin{equation}\label{dephase}
\mathcal{L}_{li}[\rho]=\sum_{n=1}^{2}L_{li}^{(n)}\rho L_{li}^{(n)^\dag}-\frac{1}{2}\{L_{li}^{(n)^\dag}L_{li}^{(n)},\rho\},
\end{equation}
where the Lindblad operators  $\mathcal{L}_{l1}$=$\sqrt{\gamma_{dp}^{1}/2}(|p\rangle\langle p|-|1\rangle\langle 1|)$ and $\mathcal{L}_{l2}$=$\sqrt{\gamma_{dp}^{2}/2}(|r\rangle\langle r|-|p\rangle\langle p|)$ describing the dephasing between $|p\rangle$ and $|1\rangle$, and between $|r\rangle$ and $|p\rangle$ caused by the phase noise of $\Omega_{1}(t)$ and $\Omega_{2}$, respectively.
Figure~\ref{Error}(a) depicts the relationship between the gate fidelity and two dephasing rates of $\gamma_{dp}^{1(2)}/2\pi\in[0,0.1]~$MHz, from which we can see that the dephasing between $|r\rangle$ and $|p\rangle$ is more influential.

(iii) The fluctuation of the detuning. The fluctuation of external magnetic field may cause a transition shift, giving a Rydberg two-photon detuning $\Delta_{B}=(g_{r}m_{r}-g_{1}m_{1})\mu_{B}B_{z}$, where $g_{r}=2$ and $g_{1}=1/2$ are Land\'{e} factors, while $m_{r}=1/2$ and $m_{1}=0$. The fluctuations of the excitation laser frequencies and the light shift will also destroy the two-photon resonance process and introduce another detuning $\Delta_{l}$. So the system Hamiltonian in this case is shown as
\begin{eqnarray}\label{Hfd}
H_{I}&=&\sum_{i=c,t}\frac{\Omega_{1}(t)}{2}|p\rangle_{i}\langle1|+\frac{\Omega_{2}}{2}|r\rangle_{i}\langle p|+{\rm H.c.}-\Delta|p\rangle_{i}\langle p|\nonumber\\&&-\frac{\delta}{2}(|1\rangle_{i}\langle 1|-|r\rangle_{i}\langle r|)+{\cal U}_{rr}|rr\rangle\langle rr|,
\end{eqnarray}
where detuning $\delta=\Delta_{B}+\Delta_{l}$. Referring to a variety of experiments,
The detuning fluctuates in accordance with a normal distribution, with a $\sigma_{\delta}$ standard deviation, on the order of a few hundred kHz. Here we choose $\sigma_{\delta}/2\pi=500~$kHz for simplicity. As shown in Table.~\ref{err}, the fluctuation of Rabi frequency has little effect on the system, and the average result after 100 repeated numerical simulations has even a negative relative error, possibly because it may compensate for the errors caused by insufficient accuracy on $T$ of the temporal pulse or the insufficiently adiabatic, etc.
\begin{figure}
\centering\scalebox{0.32}{\includegraphics{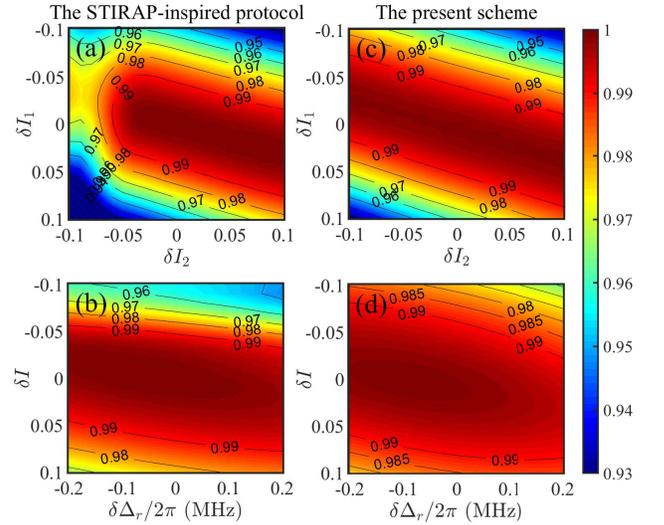}}
\caption{\label{compare}The sensitivity of the fidelity for the CZ gate to variation in optical intensity and detuning under different protocols. (a) and (b) correspond to the scheme with parameters of Fig.~8 in Ref.~\cite{PhysRevA.101.062309}. (c) and (d) corresponds to the present scheme with parameters shown in Fig.~\ref{Evolution}.}
\end{figure}
\subsection{Detection errors}
According to Ref.~\cite{PhysRevA.101.043421}, the detection errors can be divided into two parts: (i) the ``false positive" errors and (ii) the ``false negative" errors. For the ``false positive" errors, it refers to erroneously inferring an atom in the ground state as excited the $|r\rangle$ state, which can be denoted as $\epsilon=P(r|1)$. This contains the motion loss of atoms when we turn off the optical traps during the CZ gate or due to the background-gas collisions. The measured value of $\epsilon$ is typical $0.01-0.03$.
For the ``false negative" errors, it can be denoted as $\epsilon'=P(1|r)$ which is introduced by the spontaneous emission from $|r\rangle$ to $|1\rangle$ before the atom escaping. These errors can be approximated by $\epsilon'=\gamma_{r}t_{\textmd{recape}}$ when $n>50$, where $t_{\textmd{recape}}=1/\gamma_{pi}$ \cite{PhysRevA.72.022347,PhysRevA.97.053803}. $\gamma_{pi}$ is the ionization rate of the Rydberg atom which is proportional to $U_{F}$ and inversely proportional to $n^3$ where $n$ is the principal quantum number. In Ref.~\cite{PhysRevA.72.022347}, it shows that the ionization rate $\gamma_{pi}$ of the Rydberg atom with $n=50$ and $U_{F}/k_B=1~$mK is about $31000/$s. As a rough approximation, we can estimate the ionization rate corresponding to other principal quantum numbers by scaling this value like
\begin{equation}
\gamma_{pi}=\frac{U_{F}}{1\mathrm{mK}}(\frac{n}{50})^{-3}(31000)/\mathrm{s}.
\end{equation}
And then the finite error $\epsilon'$ can be estimated as $\epsilon'\approx0.0047$ including the effective lifetime of Rydberg state $100S_{1/2}$ and the measured atomic temperature. To numerically measure the gate fidelity, the state measured at the final time can be denoted as $|\psi_{t}\rangle_{m}=\alpha |00\rangle+\beta|01\rangle+\zeta|10\rangle+\eta e^{i\phi(t)}|11\rangle$, where
\begin{equation}
|\eta|^2=(1-\epsilon)^{2}\tilde{P}_{11}+(1-\epsilon)\epsilon'\tilde{P}_{1r}+\epsilon'(1-\epsilon)\tilde{P}_{r1}+\epsilon'^{2}\tilde{P}_{rr},
\end{equation}
\begin{equation}
|\zeta|^2=(1-\epsilon)\tilde{P}_{10}+\epsilon'\tilde{P}_{r0},
\end{equation}
\begin{equation}
|\beta|^2=(1-\epsilon)\tilde{P}_{01}+\epsilon'\tilde{P}_{0r}.
\end{equation}
$\tilde{P}_{jk}$ represents the population of state $|jk\rangle$ numerically. After numerical simulations, we plot the actual and corrected populations of states $|11\rangle$, $|10\rangle$, $|01\rangle$ and $|00\rangle$, respectively. The initial state is taken as $|\Psi(0)\rangle$. As shown in Fig.~\ref{Error}(b), the dotted lines are corresponding to the measured results. With $(\epsilon,\epsilon')=(0.03,0.0047)$, the detection error on fidelity is about $0.03$ while with $(\epsilon,\epsilon')=(0.01,0.0047)$ the detection error on fidelity is about $0.01$. However, these errors can be reduced by improving the detection method, such as applying strong electric field, increasing the measuring speed and improving vacuum conditions \cite{PhysRevA.97.053803,PhysRevLett.123.230501}.

In Table.~\ref{err}, we summarizes the gate errors under different technical imperfections. Among them, dephasing caused by laser phase noise has the greatest influence. And the fluctuations of laser intensity and detuning have the smallest influence even increasing the fidelity a little because of the randomness. After correcting the detection errors, the predicted gate fidelity in the experiment can reach about $0.984$ in our scheme.

\subsection{Comparison with other works in the literature}

In this section, we compare the present scheme with other previous works in terms of errors caused by the variations in the detuning and optical intensity, which can be thought to be the result of all the imperfections of the experiments. In comparison with the standard protocol that uses constant-amplitude pulses \cite{PhysRevLett.85.2208}, the application of ``STIRAP-inspired" pulse sequence in Ref.~\cite{PhysRevA.101.062309} successfully reduces the detuning sensitivity, but increases the sensitivity to intensity noise by about twice.
In Fig.~\ref{compare}(a), we reexamine the influence of the intensity noise under the corresponding parameters in figure~8 in the literature by simultaneously considering the fluctuations of both Rabi frequencies as $|\Omega_{i}(t)|^2=|\Omega_{i}(t)|^2(1+\delta I_{i})$, and in Fig.~\ref{compare}(b) we reproduce the influence of the variation on the two-photon detuning by considering a small error $\delta \Delta_{r}$, while the intensity noise is set as $\delta I_{1}=\delta I_{2}=\delta I$.

In contrast, considering the same intensity fluctuation range such as $|\Omega_{i}(t)|^2=|\Omega_{i}(t)|^2(1+\delta I_{i})$,
the combination of the adiabatic evolution and the single temporal-modulated pulse in the present scheme weaken the influence of the intensity noise, which is easy to be checked from Fig.~\ref{compare}(c). Through the simultaneous study of the variations of two-photon detuning and the light intensity on the fidelity of the scheme in  Fig.~\ref{compare}(d), we find
the present scheme is about two times less sensitivity to the intensity noise but about twice higher sensitivity to detuning compared than the protocol with ``STIRAP-inspired" pulse sequence. In this sense, our gate protocol can be considered as a compromise between the standard protocol~\cite{PhysRevLett.85.2208} and the ``STIRAP-inspired" protocol~\cite{PhysRevA.101.062309}.



\subsection{Global addressing}
\begin{figure}
\centering
\includegraphics[scale=0.25]{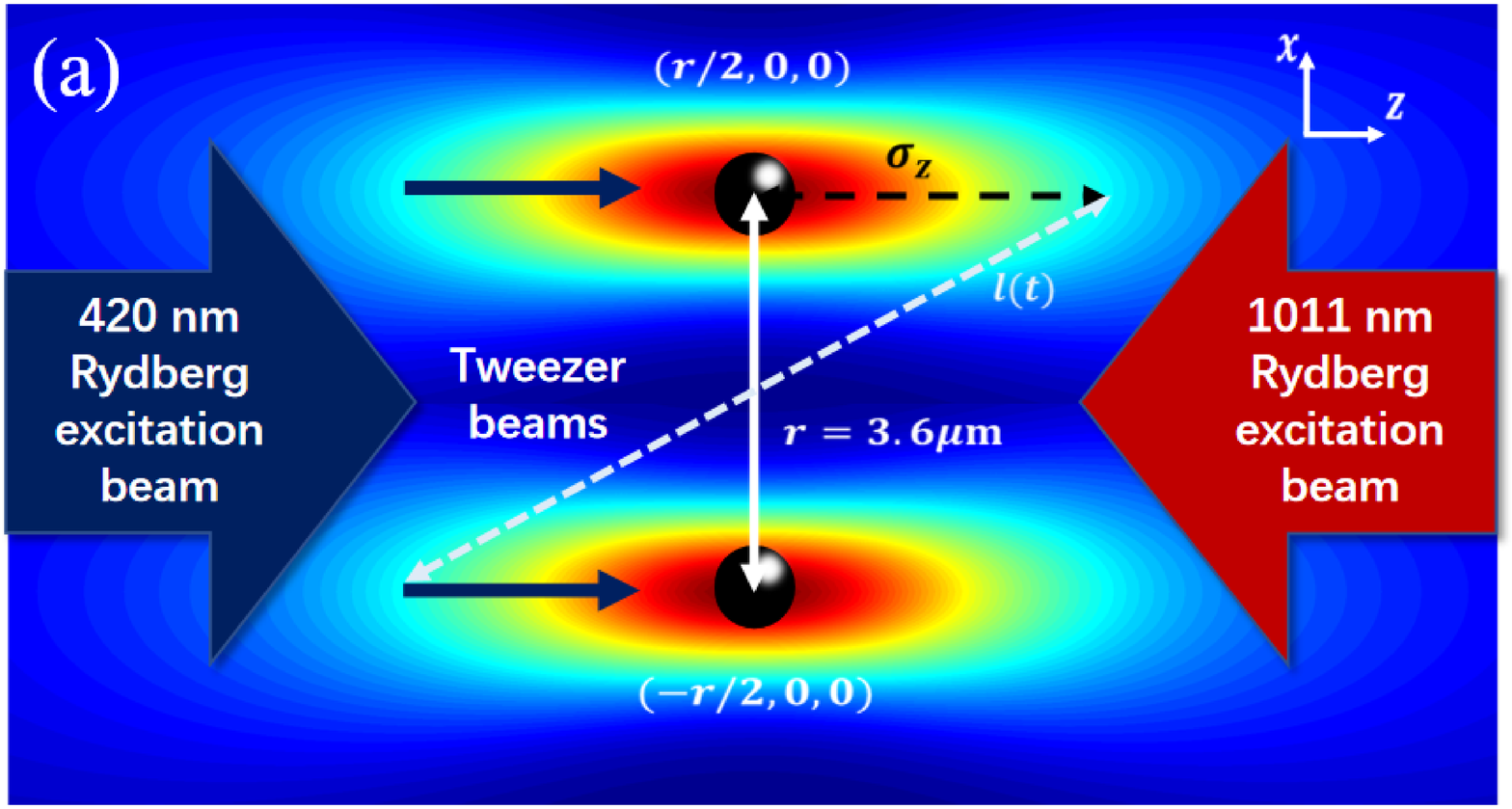}
\hspace{1in}
\includegraphics[scale=0.31]{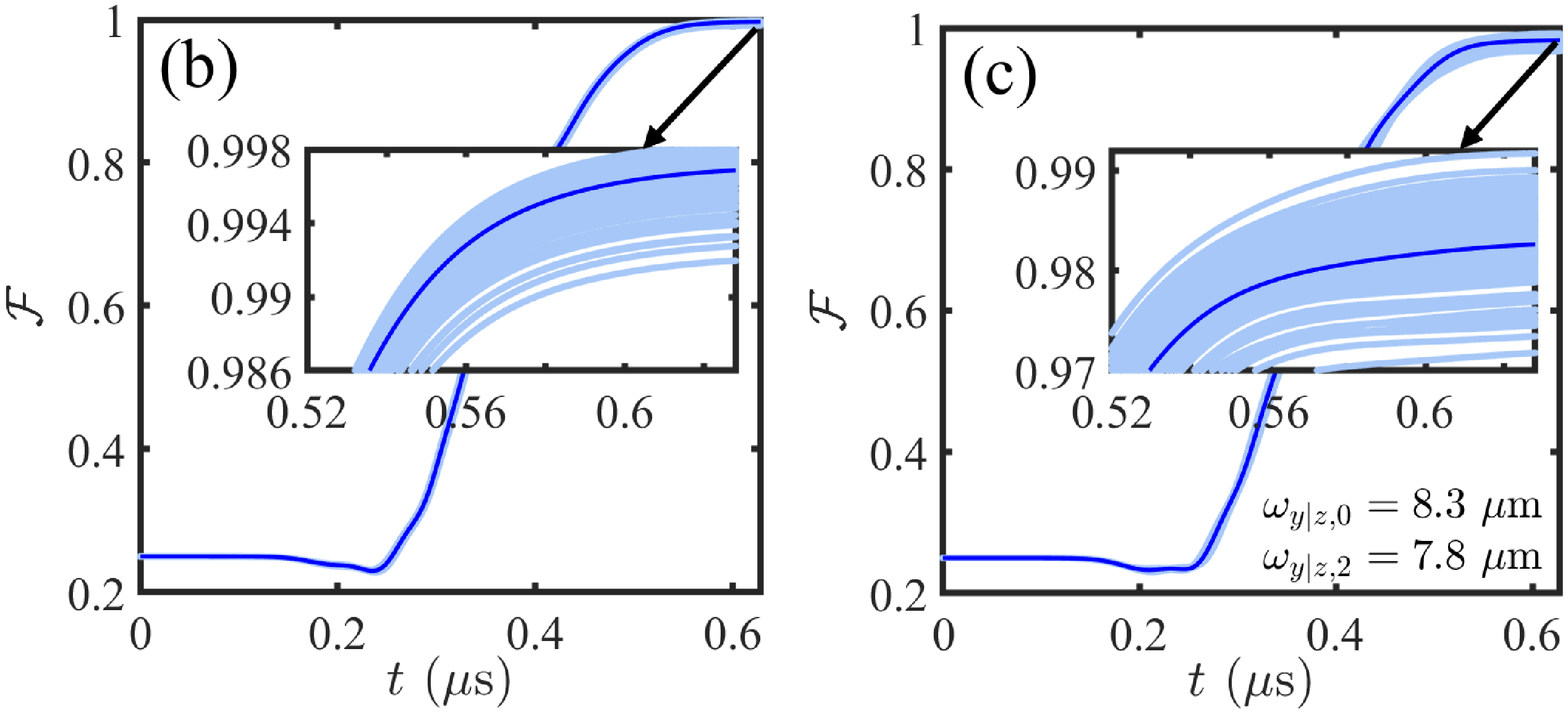}
\caption{\label{Collect1}(a) Experiment geometry. Two single atoms are trapped in two tweezers separated by about 3.6~$\mu$m on $x$ direction with tweezer beam propagating along $z$-axis. The global driving beams with waists $\omega_{x|y,0}=8.3~\mu$m and $\omega_{x|y,2}=7.8~\mu$m are counter-propagating along $z$-axis. (b) The system dynamics incorporating the Doppler effect and the fluctuation of vdW interaction at the finite temperature $T_a=5.2~\mu$K. (c) The system dynamics including the inhomogeneous Rabi frequency.}
\end{figure}
\begin{figure}
\centering
\includegraphics[scale=0.25]{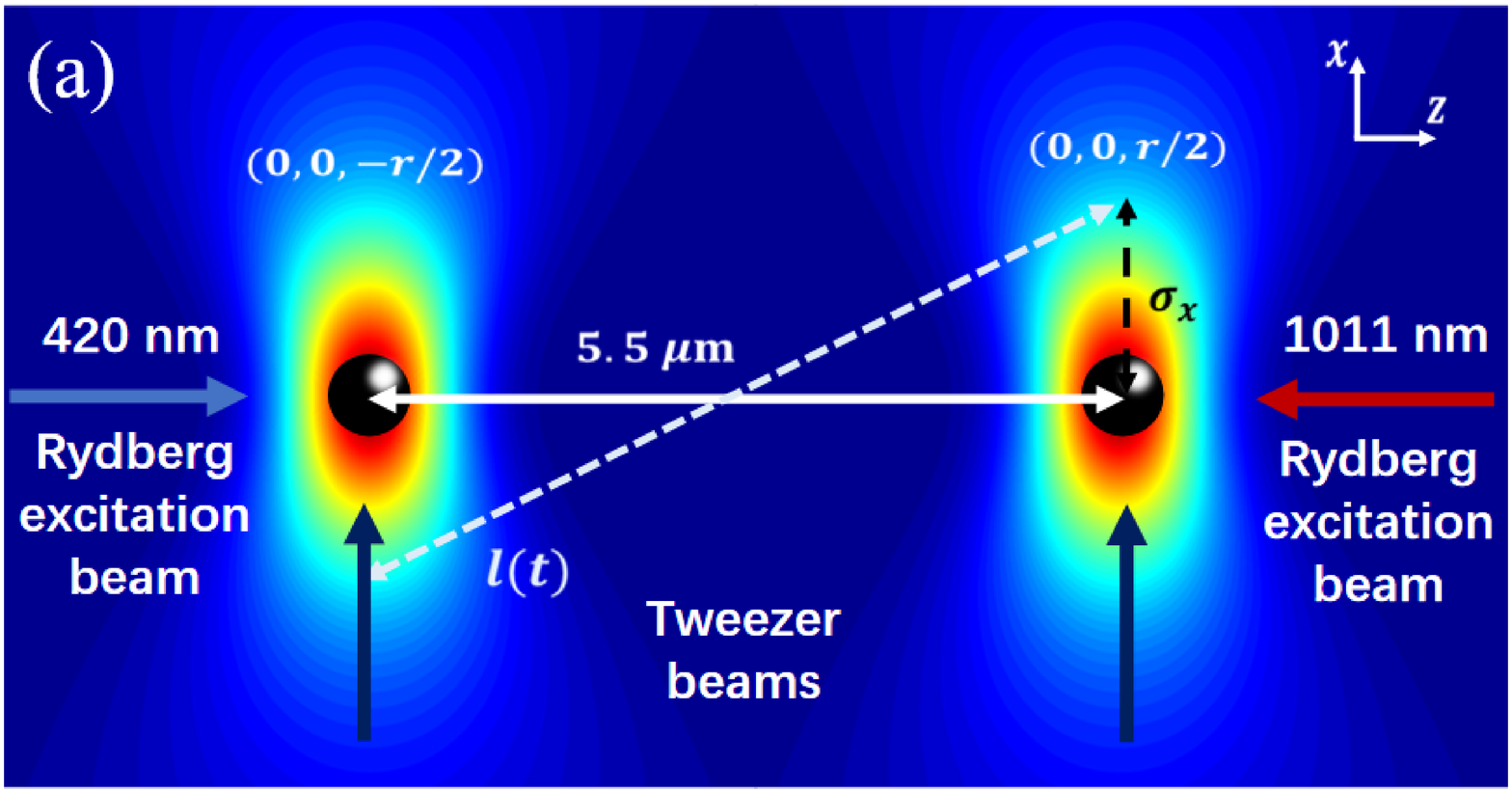}
\hspace{1in}
\includegraphics[scale=0.32]{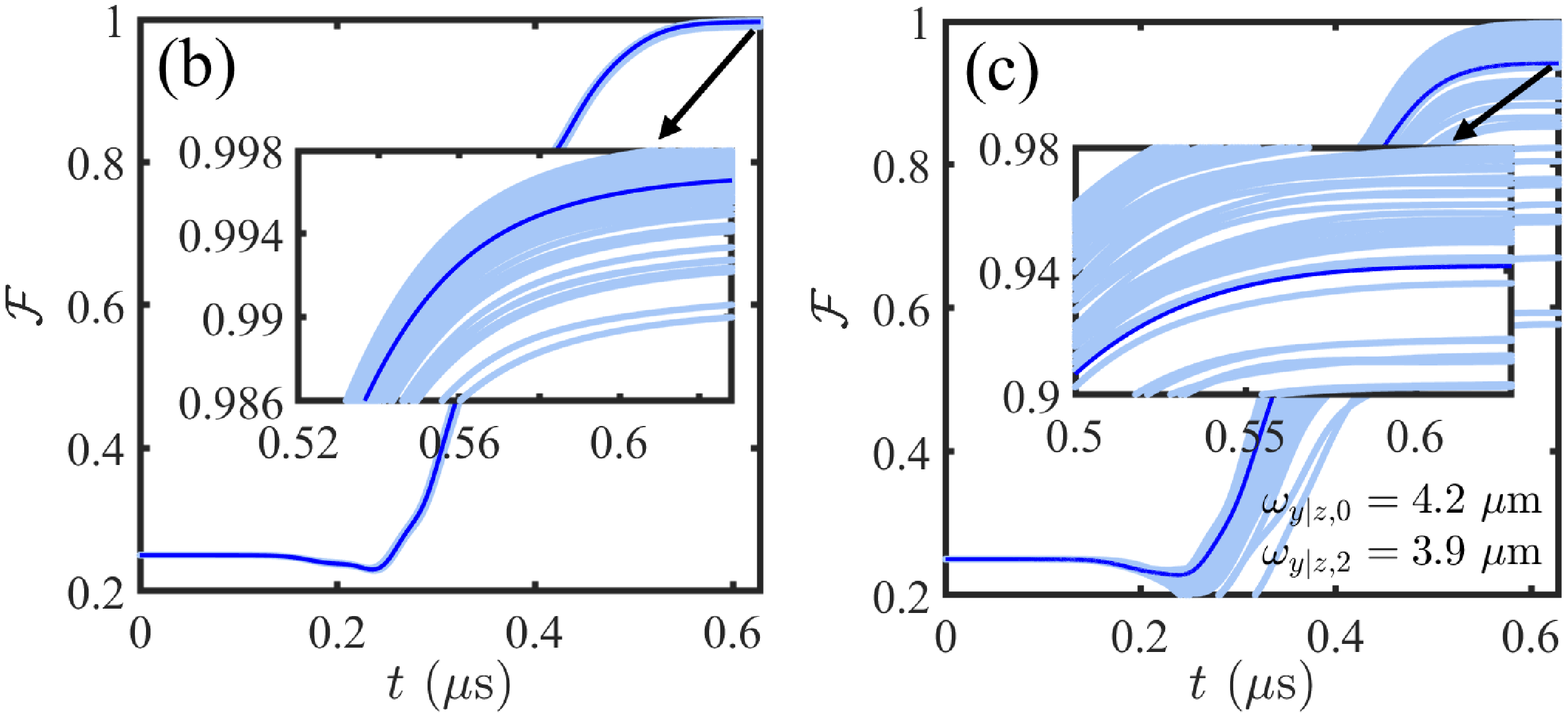}
\caption{\label{Collect2}(a) Experiment geometry. Two single atoms are trapped in two tweezers separated by about 5.5~$\mu$m on $z$ direction with tweezer beam propagating along $x$-axis. The global driving beams with waists $\omega_{x|y,0}=4.2~\mu$m and $\omega_{x|y,2}=3.9~\mu$m are counter-propagating along quantized $z$-axis, (b) The system dynamics incorporating  the Doppler effect and the fluctuation of vdW interaction at the finite temperature $T_a=5.2~\mu$K. (c) The system dynamics including the inhomogeneous Rabi frequency.}
\end{figure}
\begin{figure*}
\centering\scalebox{0.33}{\includegraphics{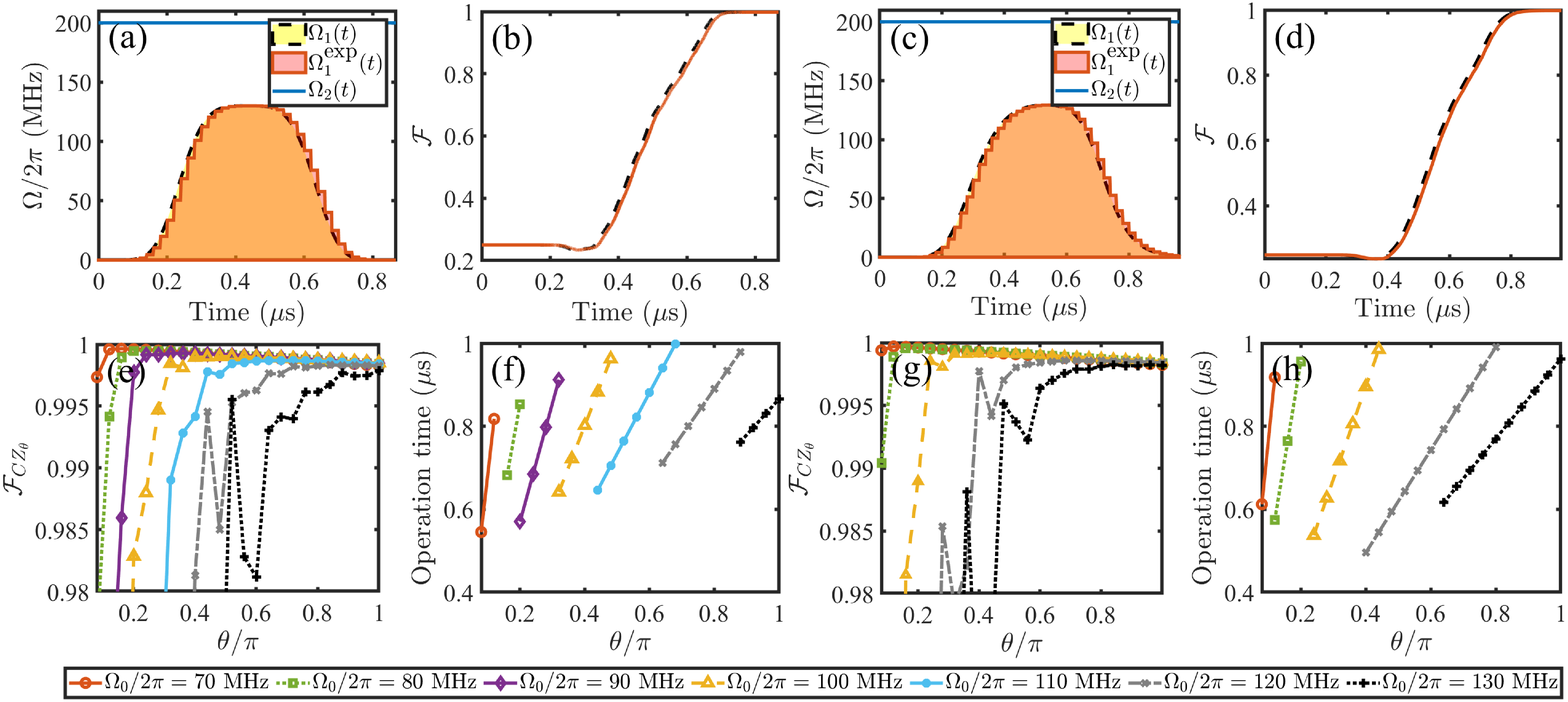}}
\caption{\label{pulses}(a) and (c) show the time dependence of Rabi frequencies of application, where $\Omega_{1}(t)$ and $\Omega_{1}^{\textmd{exp}}(t)$ correspond to the standard and experimental ones, respectively. (b) and (d) show the fidelity of the CZ gate corresponds to the above pulses, where the dotted lines correspond to the standard pulses and the solid lines correspond to the experimentally available pulses. (e)-(h) The fidelity and the total evolution time of CZ$_{\theta}$ gate with different $\Omega_{0}$ under the standard pulses shown in figures (a) and (c), respectively.}
\end{figure*}
For the convenience of experimental operation, the atoms in our scheme can also be globally driven by widening the waists of Rydberg excitation beams relative to the small interatomic spacing. As shown in Fig.~\ref{Collect1}(a), the separation of atoms is changed to $3.6~\mu$m, and the corresponding collective driving can be exploited by two lasers with waist of $8.3~\mu$m and $7.8~\mu$m \cite{PhysRevA.105.042430}. In order to achieve the laser amplitude required
previously $\{\Omega_{0}/2\pi, \Omega_{2}/2\pi\}=\{160, 200\}~$MHz, we should reset the laser beams with power $\{P_{0}, P_{2}\}$=$\{0.307\times10^{-3}, 1.162\}~$W and waist of $\{\omega_{x|y,0}, \omega_{x|y,2}\}=\{8.3, 7.8\}~\mu$m. With this updated arrangement of atoms, the error caused by atomic motion, such as Doppler shifts and the inhomogeneous Rabi frequency may change. Figs.~\ref{Collect1}(b) and \ref{Collect1}(c) respectively measures the influence of the above two kinds of experimental errors on the gate fidelity. Note that we have considered the relative position here and set $\mathbf{R}_{c,t}=(\pm r/2,0,0)$. Compared with the individual addressing scheme, the influence brought by the inhomogeneous Rabi frequency is slightly greater, because the Rabi frequencies shared with both atoms in the desired position becomes weaker.


To avoid the high laser power required for the wider beam waist, another collective driving scheme as shown in Fig.~\ref{Collect2}(a) can also be put to use.
By changing the direction of the optical tweezers and rearranging the atoms along $z$-axis perpendicular to the trap direction, the global addressing can be achieved without increasing the beam waists \cite{PhysRevA.89.033416}. Here we have set $\mathbf{R}_{c,t}=(0,0,\pm r/2)$. Comparing Figs.~\ref{Collect2}(b) with \ref{Collect1}(b), we see the errors caused by Doppler shifts in both collective driving schemes are quite similar,
but the second scheme is relatively sensitive to the inhomogeneous Rabi frequency due to the larger fluctuation range of the Rabi frequency, as indicated by Fig~\ref{Collect2}(c).
The widening fluctuation range arises from the fact that the driving pulses propagate along the $z$-axis with vibration of atoms in $x$ and $y$ directions, where the $x$-axis is also direction of the trap, which means the time average variance of the atomic position in the $x$ direction is the largest.


\section{Other forms of temporal pulses}\label{sec4}

As we mentioned in Sec.~\ref{sec2}, there are many options for the time-dependent modulation pulse as long as it meets the adiabatic conditions given by Eqs.~(\ref{ad1}) and (\ref{ad2}). To verify this point of view, we give two non-Gaussian pulses and measure the fidelity of the CZ gate. As shown in Figs.~\ref{pulses}(a) and \ref{pulses}(c), the corresponding pulse forms are a super-Gaussian pulse
\begin{equation}
(a)~~\Omega_{1}(t)=\Omega_{0}e^{-(t-2T_{1})^4/T_{1}^4}
\end{equation}
with parameters $\Omega_{0}/2\pi=130~$MHz, $\Omega_{2}/2\pi=200~$MHz and the total evolution time is $t_{tot}=4T_{1}$, and
\begin{equation}
(c)~~\Omega_{1}(t)=\Omega_{0}F_{t}\cos{(\frac{\pi}{2}f_t)},
\end{equation}
where $F_{t}=e^{-(t-T_{2})^6/2\sigma^6}$, $f_{t}=(1+e^{-4(t-T_2)/\sigma})^{-1}$ and $\sigma=0.3T_{2}$, $\Omega_{0}/2\pi=130~$MHz, $\Omega_{2}/2\pi=200~$MHz, and $t_{tot}=1.4T_{2}$.

Figs.~\ref{pulses}(b) and \ref{pulses}(d) depict the system dynamics under the standard pulses shown in figures \ref{pulses}(a) and \ref{pulses}(c) with $T_{1,2}=\{0.217,0.6875\}~\mu$s, respectively. The corresponding fidelity of the CZ gate can reach 0.9980 and 0.9982. Similarly, to be more realistic, we further measure the system dynamics under the experimentally available pulses $\Omega^{\textmd{exp}}(t)$ constructed in the same way as the experimental pulse shown in Fig.~\ref{Evolution}(a), and obtain the same gate fidelity. Besides, the advantage of an adiabatic scheme combined with a single time-dependent pulse is retained, i.e. a continuous controlled-phase gate set can be realized by considering the trade-off between the Rabi frequency and the pulse duration, as shown in Fig.~\ref{pulses}(e)-(h).

\section{Generalization and conclusion}\label{sec5}
To make a comparison with previous schemes \cite{PhysRevA.89.030301,PhysRevA.101.062309}, we further measure the fidelity of the CZ gate with $^{133}$Cs atoms.
We choose the $6S_{1/2}$ hyperfine clock states as ground states $|0\rangle\equiv|F=3, m_{F}=0\rangle$, $|1\rangle\equiv|F=4, m_{F}=0\rangle$ and the Rydberg state $|r\rangle\equiv|126S_{1/2}, m_{j}=1/2\rangle$ for concreteness.
By using a two-photon transition with $\sigma_{+}$ polarized $459~$nm and $\pi$ polarized $1038~$nm beams, the coherent Rydberg excitation between $|1\rangle$ and $|r\rangle$ can be realized where the intermediate state is chosen as $|p\rangle\equiv|7p_{1/2}, F=3, m_{F}=1\rangle$.
The lifetime of state $|p\rangle$ and $|r\rangle$ are $\tau_{p}=0.155~\mu$s and $\tau_{r}=592~\mu$s under the room temperature ($300~$K). The branching ratios equal to $b_{0(1)p}=1/16$, $b_{dp}=7/8$, $d_{1(0)r}=1/32$, $d_{dr}=7/16$, $d_{pr}=1/2$.
In such a structure, we numerically simulated the gate fidelity under the same parameters with $^{87}$Rb, i.e. $\{\Omega_{0}, \Omega_{2}\}/2\pi=\{160, 200\}~$MHz. According to the relevant levels of $^{133}$Cs, a $459$ nm beam with typical beam power $P_{0}=402~\mu$W and waist of $\omega_{x|y,0}=4~\mu$m can provide the Rabi frequency $\Omega_{0}/2\pi=160~$MHz of $|1\rangle\rightarrow|p\rangle$ transition. By tuning the $1038~$nm beam with typical beam power $P_{2}=369~$mW and waist of $\omega_{x|y,2}=2~\mu$m, we can obtain $\Omega_{2}/2\pi=200~$MHz.
The gate fidelity can reach $\mathcal{F}_{t}=0.9981$ with evolution time $T_{g}=0.628~\mu$s. Compared with the method provided in Ref.~\cite{PhysRevA.101.062309}, we obtain a higher fidelity with analytical forms of the laser pulse instead of numerical ones.



In conclusion, we have studied a method for robustly implementing a continuous controlled-phase gate set based on adiabatic evolution in the Rydberg blockade regime.
The neutral atoms are resonantly excited to Rydberg levels by a single-temporal-modulated pulse sequence individually.
According to the different adiabatic paths, a dynamical phase factor of CZ$_\theta$ gate can be accumulated on logic qubit state $|11\rangle$ alone, which can be adjusted from $0.08\pi$ to $\pi$ by calibrating the shape of the temporal pulse.
In the presence of spontaneous emission from intermediate and Rydberg states, the fidelity of CZ$_\theta$ gate can reach over 99.7\% less than 1~$\mu$s.
It is important to note that there is a wide variety of time-modulated pulse shapes in this system. It can be a Gaussian pulse or any other pulse that satisfy the adiabatic conditions, and no strict zero amplitude is required at the beginning and end.

Taking standard CZ gate as an example, we further evaluate the feasibility of the scheme from the perspective of experiment.
Using $^{87}$Rb to construct the system, the fidelity of the standard CZ gate can reach 99.78\%.
Considering various technical imperfections in the experiment, the error estimation of the CZ gate in $^{87}$Rb atomic system is discussed. Among them, the most obvious error source is the dephasing caused by laser phase noise. After correcting the detection error, the predicted fidelity can be maintained at about 98.4\%.
In addition, the global driving method is also studied, in which the influence of the inhomogeneous Rabi frequency caused by atomic vibration is more obvious. Compared with previous works in the literature, the present scheme can be considered as a compromise between the standard protocol~\cite{PhysRevLett.85.2208} and the ``STIRAP-inspired" protocol~\cite{PhysRevA.101.062309}.
In short, our gated protocol provides a robust and flexible method for adjusting the entangled phase.
We believe that this study would contribute to the experimental realization of quantum computation and quantum algorithm in the near-term neutral-atom system.


\section*{acknowledgment}
This work is supported by National Natural Science Foundation of China (NSFC) under Grants No. 11774047 and No. 12174048. W.L. acknowledges
support from the EPSRC through Grant No. EP/R04340X/1
via the QuantERA project ``ERyQSenS," the Royal Society
Grant No. IEC$\verb|\|$NSFC$\verb|\|$181078.

\bibliography{czgate}
\end{document}